\documentclass[journal,twoside,web]{ieeecolor}
\usepackage[utf8]{inputenc}
\usepackage{generic}
\usepackage{cite}
  \usepackage{amsmath,amssymb,amsfonts}
\usepackage{algorithmic}
\usepackage{graphicx}
\usepackage{subcaption}
\usepackage{algorithm,algorithmic}
\usepackage{hyperref}
\hypersetup{hidelinks=true}
\usepackage{textcomp}
\usepackage{array}
\usepackage{stfloats}
\usepackage{url}
\usepackage{verbatim}
\usepackage{cite}
\usepackage{bbm}
\usepackage{amsthm}
\usepackage[normalem]{ulem}
\usepackage{multirow}
\usepackage{mathtools}
\usepackage{booktabs}
\usepackage{makecell}

\newtheorem{Proposition}{Proposition}
\newtheorem{Theorem}{Theorem}

\newtheorem{Assumption}{Assumption}
\usepackage[justification=centering]{caption}
\usepackage{enumerate}
\def\BibTeX{{\rm B\kern-.05em{\sc i\kern-.025em b}\kern-.08em
    T\kern-.1667em\lower.7ex\hbox{E}\kern-.125emX}}
\begin{document}
\title{Probe-and-Release Coordination of Platoons at Highway Bottlenecks with Unknown Parameters}

\author{Yi Gao, Xi Xiong, Karl H. Johansson and Li Jin
\thanks{This work was in part supported by the National Natural Science Foundation of China under Grant 62103260, SJTU Global College, and J. Wu \& J. Sun Endowment Fund.}
\thanks{Y. Gao and L. Jin are with the Global College, Shanghai Jiao Tong University, China. X. Xiong is with the Key Laboratory of Road and Traffic Engineering, Ministry of Education, Tongji University, China. Karl H. Johansson is with the School of Electrical Engineering and Computer Science, KTH Royal Institute of Technology, Sweden. He is also affiliated with Digital Futures (emails: yi.gao@sjtu.edu.cn, li.jin@sjtu.edu.cn).}}

\maketitle

\begin{abstract}
This paper considers coordination of platoons of connected and autonomous vehicles (CAVs) at mixed-autonomy bottlenecks in the face of three practically important factors, viz. time-varying traffic demand, random CAV platoon sizes, and capacity breakdowns. Platoon coordination is essential to smoothen the interaction between CAV platoons and non-CAV traffic. Based on a fluid queuing model, we develop a ``probe-and-release'' algorithm that simultaneously estimates environmental parameters and coordinates CAV platoons for traffic stabilization. We show that this algorithm ensures bounded estimation errors and bounded traffic queues. The proof builds on a Lyapunov function that jointly penalizes estimation errors and traffic queues and a drift argument for an embedded Markov process. We validate the proposed algorithm in a standard micro-simulation environment and compare against a representative deep reinforcement learning method in terms of control performance and computational efficiency.
\end{abstract}


\section{Introduction}
\label{sec_introduction}

\subsection{Motivation and background}
\begin{figure}[htbp]
\centering
\begin{subfigure}{0.4\textwidth}
\includegraphics[width=\linewidth]{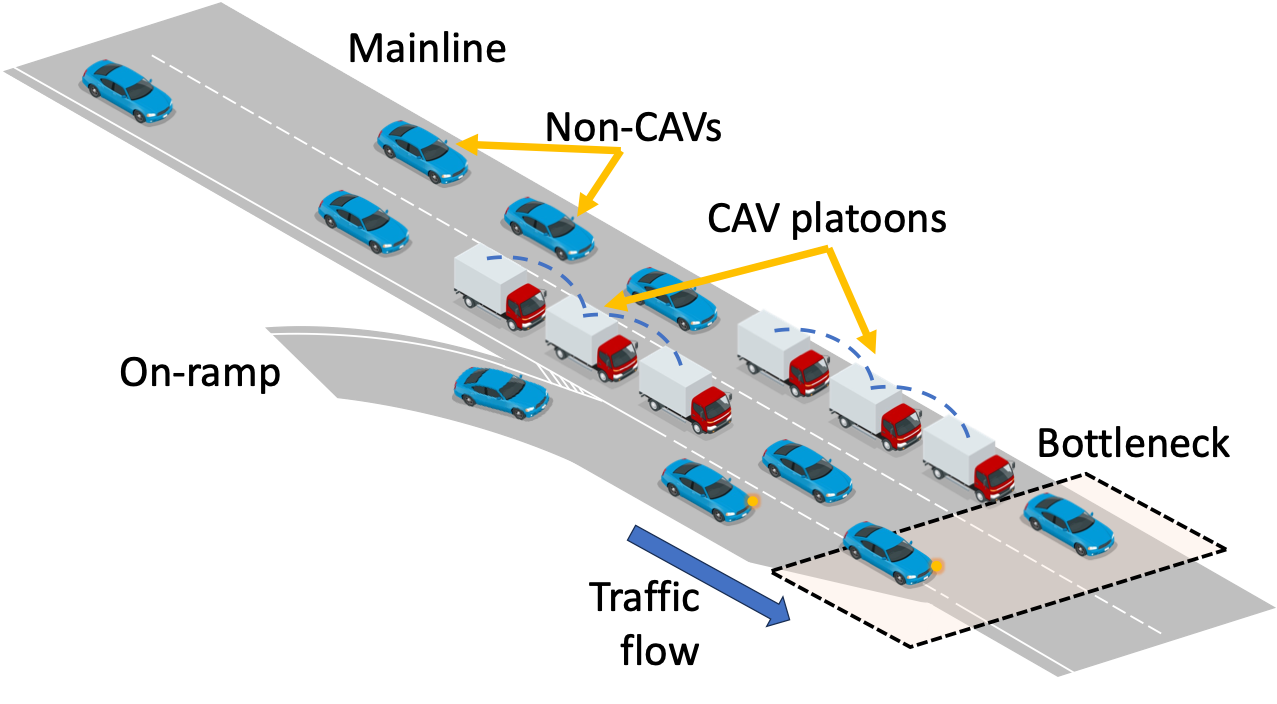}
\caption{Without coordination, CAV platoons may cause temporary congestion at the bottleneck.}
\end{subfigure}
\begin{subfigure}{0.4\textwidth}
\includegraphics[width=\linewidth]{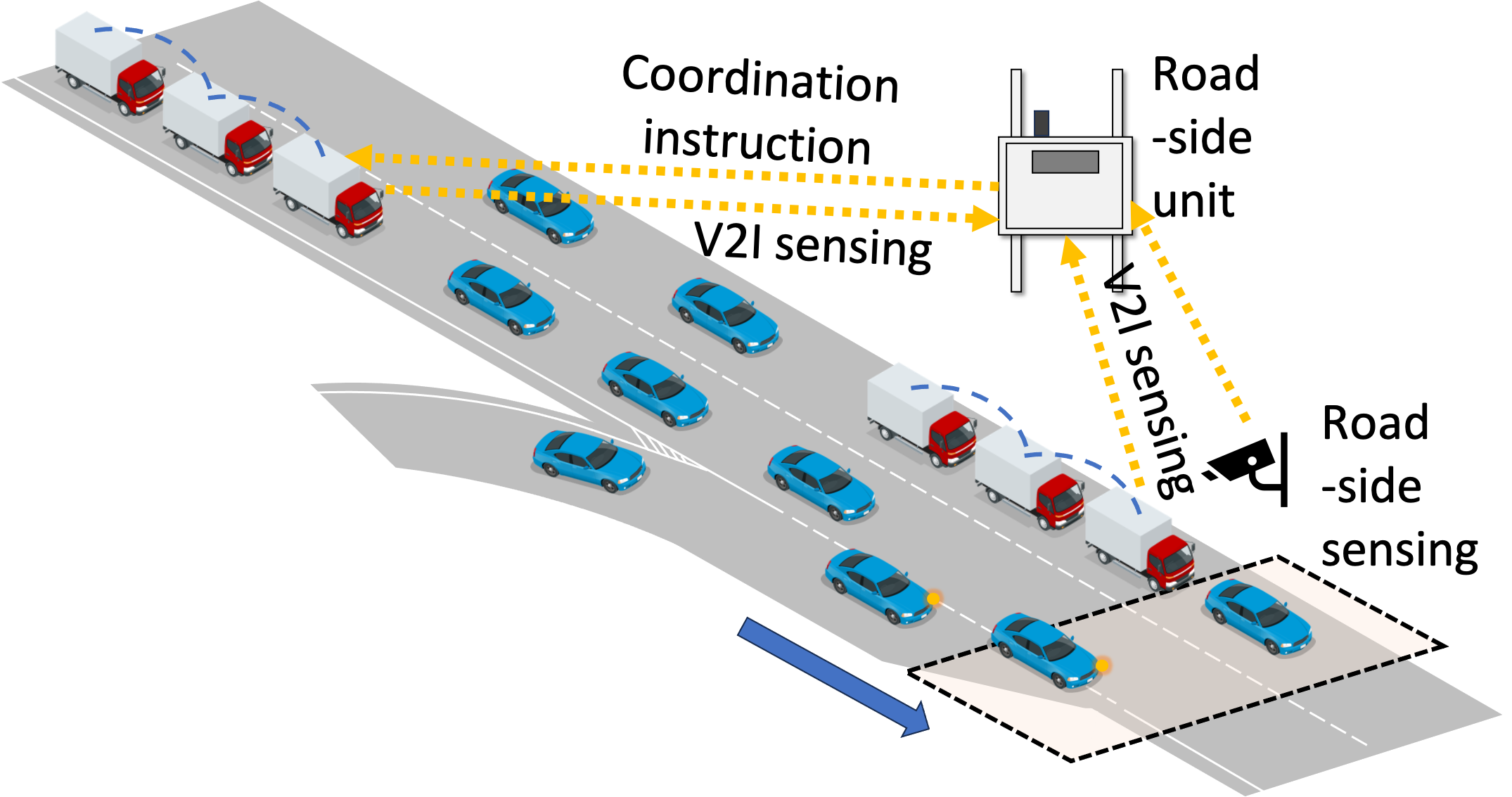}
\caption{With coordination, CAV platoons will not cluster and not disturb traffic at the bottleneck.}
\end{subfigure}
\caption{A typical mixed traffic setting with vs. without inter-platoon coordination at the bottleneck.}
\label{fig_bottleneck}
\end{figure}

\IEEEPARstart{T}{he} emerging technology of connected and autonomous vehicles (CAVs) has a strong potential for socio-economical benefits \cite{litman2017autonomous,lioris2017platoons} and is changing the practice of traffic control profoundly \cite{stern2018dissipation,zheng2020smoothing,vinitsky2023optimizing}. Since the realization of complete road traffic autonomy is currently unattainable, as acknowledged by both the academia and the industry \cite{Liu2021Computing, yurtsever2020survey,kosuru2023advancements}, for the foreseeable future, mixed traffic scenarios involving the coexistence of CAVs with non-CAVs will persist. Hence, the interaction between CAV platoons and non-CAV traffic has to be carefully managed to fully realize the benefits of platooning\cite{0Analytical,9180001,9760049}; see Fig.~\ref{fig_bottleneck}. 
However, despite extensive research achievements at the vehicle-level\cite{gao2019reinforcement,10530419,LI2025112169}, results developed for flow-level control in mixed autonomy remain relatively limited \cite{10786916,2025From}.

Another major challenge in practice is that key traffic parameters, including traffic demand, free-flow speed, capacity (saturation rate), and critical density, are unknown and/or non-stationary. These parameters are heavily influenced by external environments \cite{kyte2001effect,agarwal2005impacts,Jiyoun2009} and by internal mechanisms such as congestion-induced capacity drop\cite{2018Optimal,vcivcic2021coordinating,2022Modeling}.
Online estimation is a suitable solution to this challenge. However, convergence of the estimation error as well as the traffic state has not been sufficiently understood in this scenario. 

This paper aims to develop an adaptive control method that deals with flow-level control in mixed-autonomy traffic with unknown parameters. We consider a ``probe-and-release'' method. That is, the traffic controller periodically alternates between probing and releasing; new traffic data are collected for parameter estimation in the probing phase, while any congestion due to probing is discharged in the releasing phase. We are particularly interested in convergence behavior, control performance, and computational efficiency of this method.

\subsection{Related work}
\begin{figure*}[htbp]
    \centering
    \includegraphics[width=\textwidth]{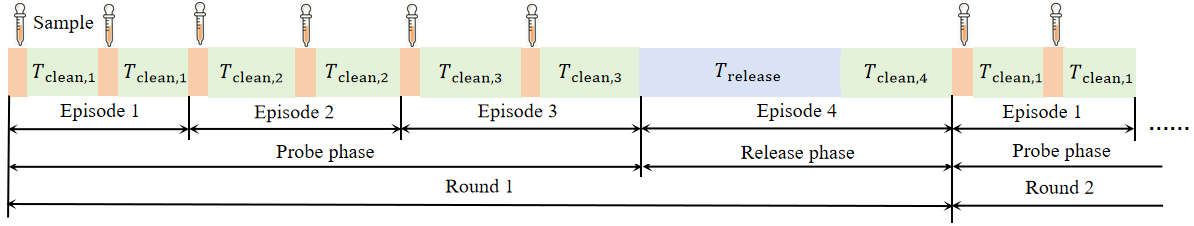}         
    \caption{The probe-and-release algorithm periodically alternates between estimation and control; a period is called a ``round''. Each episode in probe phases collects 2 samples in this illustration.}
    \label{fig_flowchart}
\end{figure*}
Platooning in terms of vehicle-level control design has been extensively studied\cite{naus2010string,coogan2014dissipativity,gao2019reinforcement,10258392}.
At the flow level, only a limited number of studies have been reported on macroscopic analysis and control of flows in partially automated traffic\cite{calvert2019evaluation, smith2020improving, zhou2021analytical,li2022cooperative,10621701}. 
In particular, Jin and \v Ci\v ci\'c et al.\cite{jin2020analysis,vcivcic2021coordinating} introduced a tandem-link fluid model of mixed-autonomy traffic flow and designed platoon coordination strategies with complete model information; the Lyapunov method in \cite{jin2020analysis} also applies to stochastic parameters. To address capacity drops, we need to resolve the non-monotonicity of the flow function. To this end, Ahmadi and Parrilo\cite{ahmadi2008non} proposed a multi-step drift method to address similar problems in deterministic settings. However, to the best of our knowledge, this technique has not been extended to the stochastic setting considered in this paper. Furthermore, these studies relied on the assumption of known environmental parameters, a condition that is difficult to satisfy in actual traffic scenarios.

A good solution to this problem is learning\cite{liu2020dynamic,peng2021dynamic,Hofleitner2012learning,han2021dynamic,chen2022data}. In particular, Vinitsky and Wu et al.\cite{vinitsky2018lagrangian,wu2021flow} convincingly demonstrated that learning-based methods have the potential to tackle unknown and non-stationary parameters. Yu and Krstic\cite{yu2022traffic} adopted a partial differential equation model for traffic dynamics and employed the learning-based method of extremum seeking to maximize flow at downstream bottlenecks with an unknown static map. 
Nevertheless, the adaptive capability of such methods in non-stationary environments has not been systematically investigated.
In addition, learning-based control methods often require a large amount of data and strong computing power, and theoretical guarantee for key performance metrics is usually limited. Meyn and Tweedie \cite{meyn2012markov} developed a generic stability theory for Markov processes, which provides a solid basis for convergence analysis of online estimation traffic control. However, ready-to-implement tools and methods have not been developed in our concrete setting.
 

\subsection{Our contributions}
First, we introduce online estimation mechanisms into the model-based design of the probe-and-release algorithm for platoon coordination in mixed autonomy.
The algorithm builds on a fluid queuing model for a representative road bottleneck with unknown parameters and with capacity drops.
We select this model to balance model realism and complexity.
Both demand (inflow) and supply (outflow) are subject to bounded white noises. 
As a baseline, we quantify the throughput loss due to the joint impact of lack of CAV coordination and of capacity drop (Proposition~\ref{prp1}).
The probe-and-release algorithm alternates between probe phases and release phases (Fig.~\ref{fig_flowchart}).
The probing phase steers the traffic states to collect effective samples for online estimation of environmental parameters. The releasing phase uses the latest estimation to coordinate CAV platoons for faster traffic release.

Second, we prove that the probe-and-release algorithm guarantees both correct estimation and stable traffic (Theorem \ref{theo_stability}).
Our analysis is grounded in the Lyapunov stability theory applicable to stochastic and/or nonlinear systems.
We ensure unbiased and convergent estimations through the basic convergence theory of stochastic processes\cite{durrett2019probability}.
To prove the stability, we construct a compound Lyapunov function coupling the estimation errors and traffic states. Our algorithm includes periodic probing and releasing and is thus non-Markovian, but the embedded process sampled at the end of each probe-and-release cycle is Markovian. We are able to apply the classical Foster-Lyapunov drift criteria \cite{meyn1993stability} to the embedded process and characterize the limiting behavior of both the estimation and the traffic state. 

Finally, we validate the probe-and-release algorithm in a realistic simulation setting and evaluate its control performance and computational efficiency. We apply the algorithm to a realistic model in Simulation of Urban Mobility (SUMO), a standard simulation tool\cite{krajzewicz2012recent}. 
We demonstrate the convergence of the algorithm. We also benchmark against  proximal policy optimization (PPO), a representative deep reinforcement learning method. The results reveal that with merely 10\% of the data required by the PPO method, our algorithm demonstrates an insignificant optimality gap around 3.96\%
in terms of the average travel time. In addition, our algorithm tracks environmental changes about 2.3 times faster than PPO.

    



The rest of this paper is organized as follows. Section \ref{sec_model} introduces the model and flow dynamics, and proposes the probe-and-release control algorithm. Section \ref{sec_Proof of main results} provides the main result and its proof. Section \ref{sec_simulation} evaluates the algorithm through simulation. Section \ref{sec_conclusion} gives concluding remarks.

\section{Model Formulation and Algorithm Design}
\label{sec_model}

In this section, we formulate the temporal queuing model for mixed-autonomy highway bottlenecks and develop the probe-and-release algorithm based on this model.

\subsection{Demand and capacity models}
\begin{figure}[htbp]
    \centering
    \includegraphics[width=0.49\textwidth]{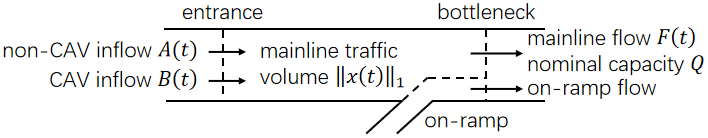}
    \caption{Highway bottleneck model; on-ramp flow is modeled as a noise to mainline flow.}
    \label{fig_abstract}
\end{figure}

Consider the representative setting of highway bottlenecks in Fig.~\ref{fig_bottleneck}, which is abstracted in Fig.~\ref{fig_abstract}.
A traffic controller observes the inflows and outflows and coordinates CAV platoons approaching the bottleneck.
The inflow consists of two classes of traffic:
\begin{enumerate}
    \item Non-CAV demand $A(t)\in\mathbb R_{\ge0}$: the number of non-CAVs entering the highway section during the $t$th time step. $A(t)$ is independent and identically distributed (i.i.d.) over time $t$ and is supported over $[A_{\mathrm{min}},A_{\mathrm{max}}]$, where $0\le A_{\mathrm{min}}<A_{\mathrm{max}}<\infty$; thus the mean $\bar{A}$ and variance $\sigma_{A}^2$ are bounded.
    
    \item CAV demand \(B(t)\in \mathbb{R}_{\geq 0}\): the size of the CAV platoon entering the highway section during the \(t\)th time step. We assume that during each time step, no more than one platoon arrives: if multiple platoons were to arrive during a very small time interval, then these platoons might have already merged into one. $B(t)$ are also i.i.d. and supported by $[0,B_{\mathrm{max}}]$, where $B_{\mathrm{max}}<\infty$; thus its mean $\bar B$ and variance $\sigma_B^2$ are also bounded.
\end{enumerate}
We assume that the controller has no prior knowledge about the statistics of demand and has to estimate in real time.

\begin{figure}[htp]
    \centering
    \includegraphics[width=0.35\textwidth]{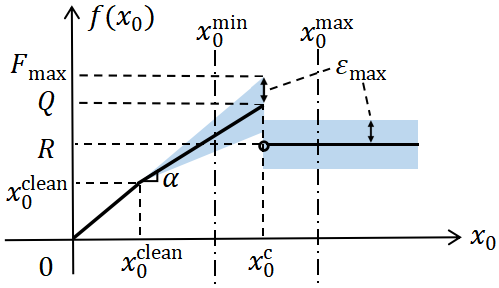}
    \caption{Flow function $f(x_0)$. Breakdown occurs at $x_0^\mathrm{c}$. Shaded area is the noise band for actual flow $F(t)$.}
    \label{fig_f}
\end{figure}

The {flow function} $f:\mathbb R_{\ge0}\to\mathbb R_{\ge0}$ 
specifies the expected number of vehicles discharged by the bottleneck per time step according to the number of vehicles queuing at the bottleneck $x_0$:
$$
f(x_0) =\left\{
\begin{array}{ll}
x_0, &0\leq x_0 \leq x_0^{\mathrm{clean}},\\
\alpha (x_0-x_0^{\mathrm{clean}})+x_0^{\mathrm{clean}}, & x_0^{\mathrm{clean}}< x_0 \leq x_0^{\mathrm{c}}, \\
R, & x_0 > x_0^{\mathrm{c}}, \\
\end{array} \right.
$$
where $0<\alpha<1$ is associated with the free-flow speed and $x_0^{\mathrm{c}}$ is the critical value, i.e., the threshold for capacity drop (Fig. \ref{fig_f}). The {nominal capacity} is given by 
$Q= \max_{x_0\geq 0} {f}(x_0) ={f}(x_0^{\mathrm{c}}),$
and the {breakdown capacity} is
$R=\lim_{x_0\to \infty} {f}(x_0).$
They illustrate that when the bottleneck is congested (i.e., $x_0>x_0^{\mathrm{c}}$), the capacity drops from the nominal value to the breakdown value. The interval $[0,x_0^{\mathrm{clean}}]$ is the clean zone, which implies that all the vehicles queuing at the bottleneck will be released if the queue length is small (i.e., $x_0\leq x_0^{\mathrm{clean}}$).
We assume that the traffic controller has no prior knowledge about $\alpha,Q,R$.
We also assume that the traffic controller has no prior knowledge about the exact value of $x_0^{\mathrm c}$ but only a range $[x_0^{\mathrm{min}},x_0^{\mathrm{max}}]$ for the value.




 


Since real traffic flow is subject to a variety of uncertainties, we assume that the actual outflow $F(t)$ is given by
$$
F(t)=\left\{
\begin{array}{ll}
f({x}_0(t)), &0\leq {x}_0(t)\leq x_0^{\mathrm{clean}},\\
f({x}_0(t))+\frac{x_0-x_0^{\mathrm{clean}}}{x_0^{\mathrm{c}}-x_0^{\mathrm{clean}}}\varepsilon(t), &x_0^{\mathrm{clean}}<{x}_0(t)\leq x_0^{\mathrm{c}},\\
f({x}_0(t))+\varepsilon(t), &{x}_0(t)>x_0^{\mathrm{c}},
\end{array} \right.
$$
where ${x}_0(t)$ is the traffic queue at the bottleneck at time step $t$, and \(\varepsilon(t)\) is an i.i.d., finite white noise with probability density function (PDF) $\varphi$ supported over $[-\varepsilon_{\mathrm{max}},\varepsilon_{\mathrm{max}}]$. Thus, its mean $\mathrm{E}[\varepsilon(t)]=0$ and variance $\sigma_{\varepsilon}^2<\infty$, as illustrated by the noise band in Fig.~\ref{fig_f}.  
We make the practical assumption that 
the uncertainty in $F(t)$ is negligible if ${x}_0(t)\leq x_0^{\mathrm{clean}}$ and gradually increases if $x_0^{\mathrm{clean}}<{x}_0(t)\leq x_0^{\mathrm{c}}$.
We also assume $\varepsilon_{\mathrm{max}}\leq (1-\alpha)(x_0^{\mathrm{c}}-x_0^{\mathrm{clean}})$ to ensure $F(t)\leq x_0(t)$, which essentially guarantees mass conservation. 
The maximum possible flow is given by $F_{\mathrm{max}}=Q+\varepsilon_{\mathrm{max}}.$
Since our focus is platoon coordination, we assume that the on-ramp is not metered and enjoys absolute priority over mainline traffic. For ease of presentation, we assume that the on-ramp inflow is purely non-CAVs. Consequently, we can capture the influence of on-ramp traffic via the noise $\varepsilon(t)$ and will not explicitly consider the on-ramp traffic.

Table \ref{table_parameters} classifies model parameters according to traffic controller's knowledge structure. The prior knowledge comes from historical data or domain expertise. The real-time observable parameters are available via modern sensing technologies. 
\begin{table}[htp]
\begin{center}
\caption{Model parameters and variables}
\label{table_parameters}
\renewcommand\arraystretch{1.5}
\begin{tabular}{
    >{\arraybackslash}m{4.5em}
    >{\arraybackslash}m{12.5em}
    >{\arraybackslash}m{4em}
    >{\arraybackslash}m{4.5em}
    } 
 \hline
Class & Quantity & Notation & Unit\\
\hline
\multirow{4}{5em}{Prior knowledge:} & free flow traverse step &$s$& step\\
&upper bound of clean zone&$x_0^{\mathrm{clean}}$& veh\\
&lower bound of critical value &$x_0^{\mathrm{min}}$& veh\\
&upper bound of critical value &$x_0^{\mathrm{max}}$&veh\\ 
\hline
 \multirow{4}{5em}{Observable variables:} & initial traffic count& ${x}_0(0)$&veh\\
 &non-CAV inflow& $A(t)$&veh/step\\ 
 &CAV inflow&$B(t)$&veh/step\\
 &actual outflow&$F(t)$&veh/step \\
 \hline
 \multirow{6}{5em}{Parameters to be estimated:} 
 &slope& $\alpha$&1/step \\
 &maximum outflow&$F_{\mathrm{max}}$&veh/step\\
 &breakdown capacity&$R$&veh/step\\ &maximum noise&$\varepsilon_{\mathrm{max}}$&veh/step\\
 &nominal capacity&$Q$&veh/step\\
 &critical value &$x_0^{\mathrm{c}}$&veh\\
 \hline
\end{tabular}
\end{center}
\end{table}

\subsection{Fluid queuing model with control}
\begin{figure}[htp]
    \centering
    \includegraphics[width=0.48\textwidth]{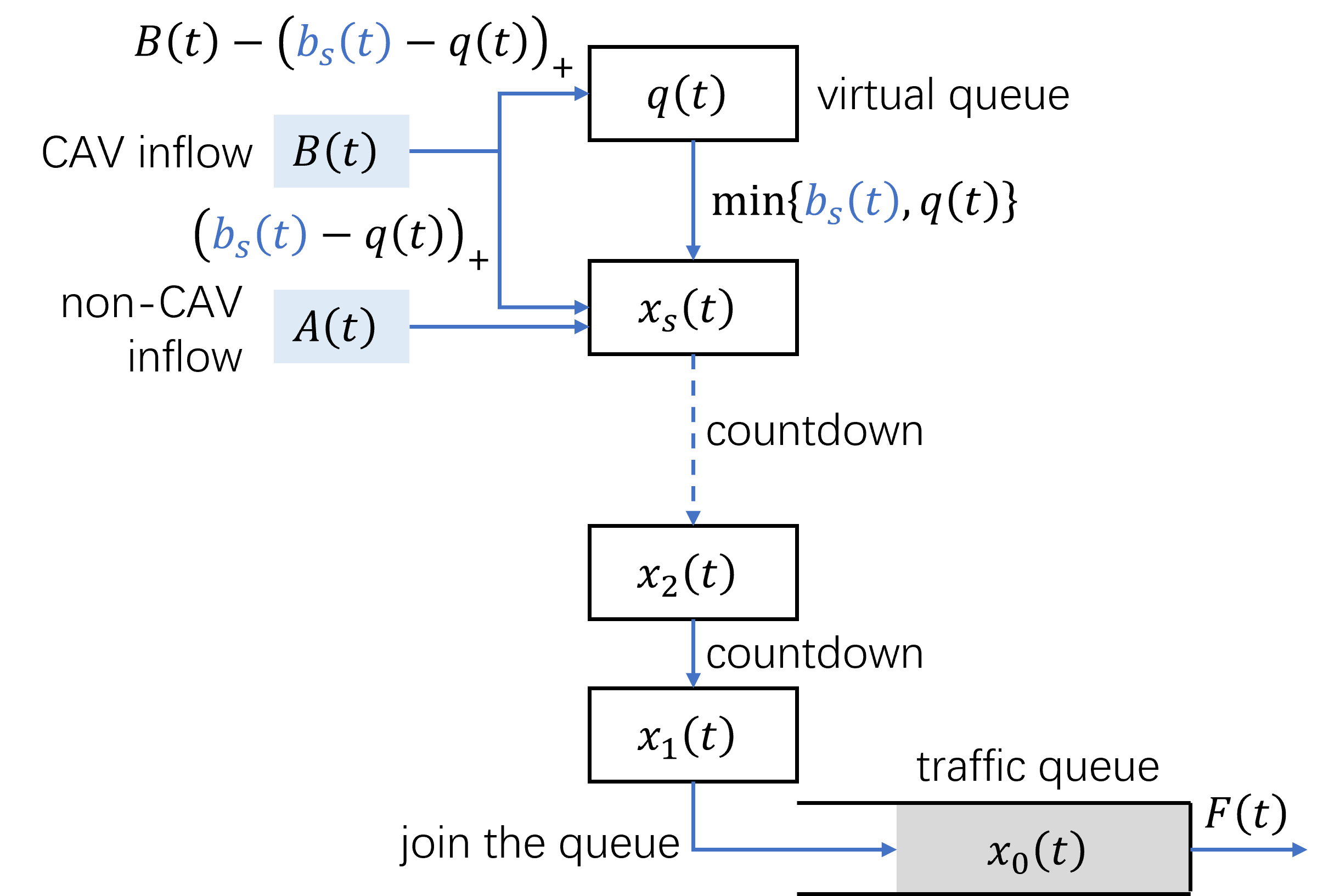}
    \caption{Fluid queuing model with delayed entrance to traffic queue. $b_s(t)$ is the only independent control input.}
    \label{fig_TCTM}
\end{figure}
We use a fluid queuing model to study traffic at the bottleneck.
Due to the spatial distance between the entrance and the bottleneck (see Fig.~\ref{fig_abstract}), we assume that the traffic $A(t)$ and $B(t)$ that enter the highway section at time step $t$ will not join the traffic queue until time step $t+s$, where $s\in\mathbb Z_{>0}$ is the free flow traverse step.
Note that $s$ is independent of traffic states and only depends on environmental information such as road type, weather and time of day\cite{agarwal2005impacts,Jiyoun2009}. To account this delay, we use ${x}_\tau(t)$ to denote the amount of traffic that will join the queue at time step $t+\tau$, where $0\le \tau\le s$; see Fig.~\ref{fig_TCTM}.

The finite-dimensional state vector of the transmission model is given by 
${x}(t)=[{x}_0(t),{x}_1(t),...,{x}_s(t),q(t)]^\mathrm{T},$
where $q(t)$ is an auxiliary state for platoon coordination to be specified below.
We initialize the state with ${x}_0(0)$ being the initial total number of vehicles on the highway section and $x_1(0)=\cdots={x}_s(0)=q(0)=0$; this does not compromise convergence analysis, since one can show that the impact of initialization error will die away after finitely many time steps.

Non-CAVs are not controlled by the traffic controller and will surely join the traffic queue after $s$ steps since entering the highway; mathematically, ${x}_s(t)$ contains all $A(t)$ non-CAVs that enter at step $t$.
The traffic controller can instruct CAV platoons to postpone their arrival to the queue. Mathematically, if the platoon $B(t)$ is not postponed, it will be contained in ${x}_s(t)$; otherwise it will be allocated to a virtual queue $q(t)$.
The control action at step $t$ is thus (i) whether the $B(t)$ CAVs entering the highway should be fully or partially allocated to $q(t)$ and (ii) whether some CAVs in the virtual queue should be released to ${x}_s(t)$.
The vector of decision variables is
$b(t)=[b_{qs}(t),b_{Bq}(t),b_{Bs}(t)]^\mathrm{T},$
where $b_{qs}(t)$ is the number of CAVs released from the virtual queue, $b_{Bq}(t)$ is the number of CAVs allocated to the virtual queue, and $b_{Bs}(t)$ is the number of CAVs that will not be postponed.
Thus, $b_{Bq}(t)=0$ for all $t$ means that no coordination is applied; this is a baseline for control performance evaluation.
Mass conservation implies that $b_{Bq}(t)+b_{Bs}(t)=B(t)$.
The total number of CAVs allocated to ${x}_s(t)$ is
$b_s(t)=b_{qs}(t)+b_{Bs}(t).$
Note that $b_{s}(t)\leq q(t)+B(t)$, which is intuitive since the desired number of CAVs should not exceed the sum of the virtual queue and the vehicles that have just arrived.
For fairness, we prioritize CAVs already in the virtual queue over newly arriving CAVs; i.e., $b_{Bs}(t)=(b_s(t)-q(t))_+$, where $(\cdot)_+$ denotes the positive part. Consequently, $b_s(t)$ is the only independent control input.
We will elaborate on how to translate the control actions above to implementable instructions to CAVs in Section~\ref{sec_simulation}.

Fig.~\ref{fig_TCTM} illustrates the update rule for the traffic state $x(t)$. Mathematically,
\begin{align*}
    &x_0(t+1)=x_0(t)+x_1(t)-F(t),\\
    &x_i(t+1)=x_{i+1}(t),\quad i=1,2,...,s-1, \\
    &x_{s}(t+1)=A(t)+b_s(t),\\
    &q(t+1)=q(t)-\min\{b_s(t),q(t)\}+B(t)-(b_s(t)-q(t))_+.
\end{align*}
Since $A(t),B(t),F(t)$ can be observed and since $b_s(t)$ is selected by the traffic controller, $x(t)$ can be uniquely determined at time $t$.
We follow \cite{meyn2012markov} and consider the system to be stable if there exists a constant $Z<\infty$ such that for any initial condition,
\begin{align}
    \limsup_{t\to\infty}\frac1t\int_{r=0}^t\mathrm{E}[\|x(r)\|_1]dr\le Z,
    \label{eq_stability}
\end{align}
where $\|x(r)\|_1$ is the total traffic count at time step $r$.
This notion of stability enables us to quantify the throughput loss due to the capacity drop. In particular, we have the following preliminary result in the absence of inter-platoon coordination, which is proved in the Appendix.
\begin{Proposition} [Baseline throughput]
\label{prp1}
    Without coordination (i.e., $b_{Bq}(t)=0$ for all $t$), the system is stable in the sense of \eqref{eq_stability} if and only if at least one of the following conditions is satisfied:
\begin{enumerate}

\item $\Bar{A}+\bar{B}< R$,

\item ${x}_0(0)\leq x_0^{\mathrm{c}}$ and $A_{\mathrm{max}}+B_{\mathrm{max}}\leq Q-\varepsilon_{\mathrm{max}}$.

\end{enumerate}
\end{Proposition}
Intuitively, condition 1) means that the total demand is less than the dropped capacity, and condition 2) means that the initial state is luckily in an invariant set without capacity drops. Both conditions are quite restrictive, so we require coordination to address such restriction.



\subsection{Probe-and-release algorithm}
\label{sec_probe_and_release}


We determine the control input with a probe-and-release algorithm. To mitigate vehicles that are continuously accumulated due to probing, the releasing phase follows probing. To track environmental parameters, the probing phase resumes after releasing. Thus, the algorithm periodically alternates between a probing phase and a releasing phase.
Let a round be a probing phase followed by a releasing phase.
Suppose that we have completed the $(n-1)$th round and are starting the $n$th round.
Suppose that $x_0(t)$ is in the ``clean'' interval $[0,x_0^{\mathrm{clean}}]$ at the end of the $(n-1)$th round; we will justify this assumption when the complete algorithm is defined.
Let $\hat\alpha[n-1],\hat F_{\max}[n-1],\hat R[n-1],\hat\varepsilon_{\max}[n-1]$ be the estimates of the model parameters at the end of the $(n-1)$th round.
The update procedures are as follows.


\subsubsection{Probing phase}
\label{sec_probe phase}

 
The objective of this phase is to obtain sample points for the flow function $f(x_0)$ depicted in Fig.~\ref{fig_f}.
The essential idea of this phase is to steer the traffic state $x_0(t)$ to the three intervals $[x_0^{\mathrm{clean}},x_0^{\min}]$, $[x_0^{\min},x_0^{\max}]$, and $[x_0^{\max},\infty)$ in three episodes sequentially; this ensures that effective samples for $(x_0(t),F(t))$ are collected in all three regimes of the flow function. The switching between different episodes is essentially
the switching between different set values $x_0^{\mathrm{set}}$ of the traffic state $x_0(t)$.  Samples in $[x_0^{\mathrm{clean}},x_0^{\min}]$ are used to estimate $\alpha$. Samples in $[x_0^{\min},x_0^{\max}]$ are used to estimate $F_{\max}$. Samples in $[x_0^{\max},\infty)$ are used to estimate $R$ and $\varepsilon_{\max}$. With the determination of the four
parameters, the flow function can be completely obtained.

\begin{Assumption}
\label{ass_2}
    There exist positive constants  $\delta_1,\delta_2,\Lambda$ known to the traffic controller such that 
    \begin{align*}
        &\delta_1 \leq \min\{x_0^{\mathrm{clean}},R-\varepsilon_{\mathrm{max}}\}-A_{\mathrm{max}},\\ &\delta_2\leq \tilde{R}-\bar{A}-\bar{B},\quad\Lambda\geq A_{\mathrm{max}}+B_{\mathrm{max}}.
    \end{align*}
\end{Assumption}

Episode 1: First, select $x_0^{\mathrm{set}}$ uniformly at random from the interval $[x_0^{\mathrm{clean}}, x_{0}^{\mathrm{min}}]$. 
Set $b_s(t)=x_0^{\mathrm{set}}(t)-A(t)$ for one step; this steers $x_0(t+s+1)$ to the interval $[x_0^{\mathrm{clean}},x_0^{\min}]$ and thus collects an effective sample $(x_0(t+s+1),F(t+s+1))$ therein.
Second, set $b_s(t)=0$ for the next $T_{\mathrm{clean},1}$ steps, where $T_{\mathrm{clean},1}=\lceil{(x_0^{\mathrm{min}}-x_0^{\mathrm{clean}})}/{\delta_1}\rceil.$
This must steer $x_0(t)$ to the clean interval after $T_{\mathrm{clean},1}$ steps; the proof is straightforward but tedious and is therefore not included.
If $k$ samples are required in an episode, the above steer-clean procedure will be done for $k$ times (Fig. \ref{fig_flowchart}), which will take $k(1+T_{\mathrm{clean},1})$ steps.
Let $(x_0(t_1),F(t_1)),(x_0(t_2),F(t_2)),\ldots,(x_0(t_k),F(t_k))$ be the $k$ samples.
Note that the cleaning makes the $k$ samples statistically independent.
Let $\vartheta_{\alpha}^{(j)}=(F(t_j)-x_0^{\mathrm{clean}})/(x_0(t_j)-x_0^{\mathrm{clean}})$, which is actually the $j$th estimate for parameter $\alpha$.
The outcome of this episode is $k$ estimates $\{\vartheta_{\alpha}^{(1)},\vartheta_{\alpha}^{(2)},\ldots,\vartheta_{\alpha}^{(k)}\}$ for $\alpha$.
The state $x_0(t)$ is also guaranteed to be cleaned at the end of this episode.

Episode 2: First, select 
$x_0^{\mathrm{set}}$  uniformly at random from $[x_0^{\mathrm{min}},x_0^{\mathrm{max}}]$.
Set $b_s(t)=x_0^{\mathrm{set}}(t)-A(t)$ for one step; since $x_0(t)$ was cleaned at the end of episode 1, this steers $x_0(t+s+1)$ to the interval $[x_0^{\mathrm{min}},x_0^{\mathrm{max}}]$.
Second, set $b_s(t)=0$ for next $T_{\mathrm{clean},2}$ steps, where
$T_{\mathrm{clean},2}=\lceil{(x_0^{\mathrm{max}}-x_0^{\mathrm{clean}})}/{\delta_1}\rceil$; this must steer $x_0(t)$ to the clean interval $[0,x_0^{\mathrm{clean}}]$.
Conducting the above procedures for $k$ times, we will have $k$ sample points $\{(x_0(t_1),F(t_1)),(x_0(t_2),F(t_2)),\ldots,(x_0(t_k),F(t_k))\}$ (with a slight abuse of notation) in the interval $[x_0^{\mathrm{min}},x_0^{\mathrm{max}}]$. 
This episode will take $k(1+T_{\mathrm{clean},2})$ steps. We set
$\vartheta_{F_{\mathrm{max}}}^{(j)}=F(t_j),j=1,2,\dots,k$.
The outcome of this episode is $k$ estimates $\{\vartheta_{F_{\mathrm{max}}}^{(1)},\vartheta_{F_{\mathrm{max}}}^{(2)},\ldots,\vartheta_{F_{\mathrm{max}}}^{(k)}\}$ for $F_{\max}$.
The state $x_0(t)$ is also guaranteed to be cleaned at the end of this episode.


Episode 3: 
First, select 
$x_0^{\mathrm{set}}$ uniformly at random from $[x_0^{\mathrm{max}},1.5x_0^{\mathrm{max}}]$.
Set $b_s(t)=x_0^{\mathrm{set}}(t)-A(t)$ for one step; since $x_0(t)$ was cleaned at the end of episode 2, this steers $x_0(t+s+1)$ to the interval $[x_0^{\mathrm{max}},1.5x_0^{\mathrm{max}}]$.
Second, set $b_s(t)=0$ for next $T_{\mathrm{clean},3}$ steps, where
$T_{\mathrm{clean},3}=\lceil{(1.5x_0^{\mathrm{max}}-x_0^{\mathrm{clean}})}/{\delta_1}\rceil$; this must steer $x_0(t)$ to the clean interval $[0,x_0^{\mathrm{clean}}]$.
Conducting the above procedures for $k$ times, we will have $k$ sample points $\{(x_0(t_1),F(t_1)),(x_0(t_2),F(t_2)),\ldots,(x_0(t_k),F(t_k))\}$ (again with a slight abuse of notation) in the interval $[x_0^{\mathrm{max}},1.5x_0^{\mathrm{max}}]$. This episode will take $k(T_{\mathrm{clean},3}+1)$ steps. Let $\vartheta_{R}^{(j)}=F(t_j)$, which is actually the $j$th estimate of $R$.
The outcome of this episode is $k$ estimates $\{\vartheta_{R}^{(1)},\vartheta_{R}^{(2)},\ldots,\vartheta_{R}^{(k)}\}$ for $R$.
The state $x_0(t)$ is also guaranteed to be cleaned at the end of this episode.

Now we are ready to update our estimates as follows:
\begin{subequations}
\begin{align}
\label{equ_alpha} &\hat{\alpha}[n]=\sum_{j=1}^{k}\lambda(1-\lambda)^{k-j} \vartheta_{\alpha}^{(j)}+(1-\lambda)^k\hat{\alpha}[n-1],\\
\label{equ_F} &\hat{F}_{\mathrm{max}}[n]=\max\Big\{\hat{F}_{\mathrm{max}}[n-1],\max_{1\le j\le k}\vartheta_{F_{\mathrm{max}}}^{(j)}\Big\},\\
\label{equ_R} &\hat{R}[n]=\sum_{j=1}^{k}\lambda(1-\lambda)^{k-j} \vartheta_{R}^{(j)}+(1-\lambda)^k\hat{R}[n-1],\\
\label{equ_noise} &\hat\varepsilon_{\mathrm{max}}[n]=\max\Big\{\hat\varepsilon_{\mathrm{max}}[n-1],\frac {1}{2}\Big(\max_{1\le j\le k}\vartheta_{R}^{(j)}-\min_{1\le j\le k}\vartheta_{R}^{(j)}\Big)\Big\}.
\end{align}
\end{subequations}
We select a constant learning rate $\lambda$ to track possibly non-stationary environments.
The critical value $\hat x_0^{\mathrm c}$ can be derived as
$$\hat x_0^{\mathrm{c}}[n]=x_0^{\mathrm{clean}}+\frac{\hat{F}_{\mathrm{max}}[n]-\hat{\varepsilon}_{\mathrm{max}}[n]-x_0^{\mathrm{clean}}}{\hat\alpha[n]}.$$
The normalized estimation errors are given by
\begin{align*}
& e_\alpha[n] = \frac{\hat{\alpha}[n]-\alpha}{\alpha}, 
\ e_{F_{\mathrm{max}}}[n] = \frac{\hat F_{\mathrm{max}}[n] - (Q+\varepsilon_{\mathrm{max}})}{Q+\varepsilon_{\mathrm{max}}},\\
& e_R [n]= \frac{\hat R[n]-R}{R},\ e_{\varepsilon_{\mathrm{max}}}[n]=\frac{\hat{\varepsilon}_{{\mathrm{max}}}[n]-\varepsilon_{\mathrm{max}}}{\varepsilon_{\mathrm{max}}}.
\end{align*}
Let $e[n]=[e_{\alpha}[n],e_{F_{\mathrm{max}}}[n], e_R[n],e_{\varepsilon_{\mathrm{max}}}[n]]^{\mathrm{T}}$. 
Due to the non-vanishing learning rate $\lambda$, the error will not converge to a constant limit.
Instead, we are interested in whether the error can be bounded from above, i.e., finding a constant $Y<\infty$ such that for any initial condition,
\begin{align}
\label{equ_def_convergence}
\limsup _{t \rightarrow \infty} \frac{1}{t} \sum_{n=0}^{t} \mathrm{E}[\|e[n]\|_2^2] \leq Y,
\end{align}
where $\|\cdot\|_2$ is the 2-norm. Note that the above update rule may not track changes (especially decreases) in $F_{\max}$ and $\varepsilon_{\max}$. To fix this, we reset $\hat F_{\max}[n]$ and $\hat\varepsilon_{\max}[n]$ to small default values with a period (in the order of hours) much longer than the update step (in the order of seconds); we will demonstrate the consequences of such resets in Section~\ref{sec_simulation}.





\subsubsection{Release phases}
The essential idea of the releasing phase is to steer $x_0(t)$ to 
the estimated critical value $\hat x_0^{\mathrm c}[n]$. The control input $b_s(t)$ is selected based on the current state $x(t)$ and the predictions for $\{x_0(t+1),\ldots,x_0(t+s)\}$, which we denote as $\{x_0^{\mathrm{pred}}(t+1),\ldots,x_0^{\mathrm{pred}}(t+s)\}$; also define $x_0^{\mathrm{pred}}(t)=x_0(t)$. The estimated flow function is
$$
\hat f^{[n]}(x_0) =\left\{
\begin{array}{ll}
x_0, \hspace{1.5cm}0\leq x_0 \leq x_0^{\mathrm{clean}},\\
\hat\alpha[n] (x_0-x_0^{\mathrm{clean}})+x_0^{\mathrm{clean}}, \\
\hspace{2cm}x_0^{\mathrm{clean}}< x_0 \leq \hat x_0^{\mathrm{c}}[n], \\
\hat R[n], \hspace{1.1cm}x_0 > \hat x_0^{\mathrm{c}}[n],
\end{array} \right.
$$
Then, for $0\leq \ell \leq s-1$, we have
\begin{equation}
   \begin{aligned}
\label{equ_x0_pred}
    {x}_0^{\mathrm{pred}}(t+\ell+1)={x}_0^{\mathrm{pred}}(t+\ell)+x_{\ell+1}(t)-&\hat f^{[n]}\big(x_0^{\mathrm{pred}}(t+\ell)\big).
\end{aligned} 
\end{equation}
Now we are ready to specify the control inputs.

Episode 4: First, let 
$x_0^{\mathrm{set}}={\hat{x}_0^{\mathrm{c}}}[n]$ and set the ideal control input to be
\begin{equation*}
\label{equ_bs_1}
\begin{aligned}
    b^{\ast}_{s}(t)=x_0^{\mathrm{set}}-{x}_0^{\mathrm{pred}}(t+s)
    +\hat{f}^{[n]}({x}_0^{\mathrm{pred}}(t+s
))
    -A(t).
\end{aligned}
\end{equation*}
Apply the control input
$
b_s(t)=\min\left\{(b_s^*(t))_+, q(t) + B(t)\right\}
$
for
\begin{equation}
\label{equ_T_rel}
T_\mathrm{release}=\bigg\lceil\frac{(\mu_1-1)\Lambda(3k+k\sum_{i=1}^{3}T_{\mathrm{clean},i}+T_{\mathrm{clean,4}})}{\Lambda+\mu_1\delta_2}\bigg\rceil
\end{equation}
steps, where $\mu_1$ is a constant in $(-\infty,\Lambda/(-\delta_2))$.
Second, apply $b_s(t)=0$ for $T_{\mathrm{clean},4}=\lceil{((s+1)x_0^{\mathrm{max}}-x_0^{\mathrm{clean}})}/{\delta_1}\rceil$ steps; the state $x_0(t)$ is also guaranteed to be cleaned thereafter. 

For the probe-and-release algorithm, the initial probing results may be bad, leading to bad initial release control. The lucky thing is that the poor release phase ends after some time steps and we can expect better performance in the following round. Both the probe phase and the release phase are guaranteed to be good eventually.

The following technical definitions will be used subsequently.
For a small positive number $\gamma$, define
\begin{equation}
\label{equ_Kappa}
\begin{aligned}
    \kappa(\gamma)=\mathrm{min}\Big\{&(1-(1-\lambda)^{2k})\gamma,\\
    &\frac{7\gamma p(\sqrt{\gamma}(Q+\varepsilon_{\mathrm{max}}))}{16},\frac{7\gamma p'(\sqrt{\gamma}\varepsilon_{\mathrm{max}})}{16}\Big\},
\end{aligned}
\end{equation}
where $p(\chi)$ and $p'(\psi)$ are auxiliary functions given in the appendix.
Let $\xi_0=[x_0^{\mathrm{clean}}+\frac{Q-x_0^{\mathrm{clean}}}{(1-2\sqrt{\gamma})\alpha},Q,\ldots,Q,(M-s-1)Q]^\mathrm{T}\in\mathbb R^{s+2}$, where
$
    M = \lceil{s(Q-R+\varepsilon_{\mathrm{max}})}/{\delta_1}\rceil+s.
$
Define
\begin{align}
\label{equ_R_tilde}
    \tilde{R}=\frac{1}{M}\sum_{m=1}^{M}\mathrm{E}[f({x}_0(m))|x(0)=\xi_0].
\end{align}
The expectation is with respect to the i.i.d. random variables.
The computation of $\tilde{R}$ is straightforward but tedious and is therefore moved to the appendix.
Note that all the above definitions are independent of real-time system states and can be computed a priori.
Also note that although these definitions depend on model information, they are not required by our control algorithm; they will be used only for convergence analysis.
\section{Stability analysis}
\label{sec_Proof of main results}
In this section, we show that the probe-and-release algorithm (i) attains a closed-form upper bound for the estimation errors and (ii) stabilizes the traffic queue despite the potential congestion due to probing actions. The main result of this paper is as follows.
\begin{Theorem}
\label{theo_stability}
Suppose that the probe-and-release algorithm is applied to the fuild queuing model under Assumption \ref{ass_2}. Then, the estimation errors are bounded in the sense of (\ref{equ_def_convergence}) with 
$$Y=\bigg(\frac{1}{R^2}+\frac{1}{\alpha^2{(x_0^{\mathrm{c}}-x_0^{\mathrm{clean}}})^2}\bigg)\frac{\lambda\sigma_{\varepsilon}^2}{2-\lambda}.$$
Furthermore, the traffic state is stable in the sense of (\ref{eq_stability}) if 
\begin{enumerate}
    \item[(i)] $\bar{A}+\bar{B}<\tilde{R}$, where $\tilde{R}$ is given by (\ref{equ_R_tilde});
    \item [(ii)] $A_{\mathrm{max}}<\min\{x_0^{\mathrm{clean}},R-\varepsilon_{\mathrm{max}}\}$;
    \item[(iii)] there exists $\gamma>0$ such that
    \begin{equation}
    \label{equ_noise_var}
        \sigma_{\varepsilon}^2< \frac{\kappa(\gamma)\delta_2(2-\lambda)\alpha^2R^2(x_0^{\mathrm{c}}-x_0^{\mathrm{clean}})^2}{\lambda(\delta_2+\Lambda)(1-(1-\lambda)^{2k})(R^2+\alpha^2(x_0^{\mathrm{c}}-x_0^{\mathrm{clean}})^2)},
    \end{equation}
where $\kappa(\gamma)$ is defined by (\ref{equ_Kappa}).
\end{enumerate}

\end{Theorem}

The theorem indicates that the estimation error bound $Y$ decreases as the variance of the system noise $\sigma_{\varepsilon}^2$ decreases, which is intuitive. The bound decreases as the learning rate $\lambda$ decreases; this actually involves the balance between convergence speed and estimation accuracy. 
The three conditions for traffic stability can be interpreted as follows. (i) The releasing phase is functional if the total inflow $\bar A+\bar B$ is upper bounded by $\tilde R$, which is an estimation of the time-average bottleneck capacity possibly in the presence of intermittent capacity drops. 
(ii) The maximum of non-CAV inflow, which cannot be controlled, should be sufficiently small so that the traffic state can be steered to any interval in Fig.~\ref{fig_f} as desired by the probe-and-release algorithm.
(iii) The releasing phase requires an upper bound for the variance of the system noise; otherwise the traffic state will be subject to excessive randomness and thus be hard to control.
A numerical case study is provided in Section~\ref{sec_simulation} to further illustrate the ideas of this theorem.

The rest of this section is devoted to the proof of the theorem. Section~\ref{sub_error} focuses on the probing phase and examines the estimation process. Section~\ref{sub_traffic} focuses on the releasing phase and establishes the stability condition.

\subsection{Convergence of estimation errors}\label{sub_error}
An important characteristic of the fluid queuing system under the probe-and-release algorithm is an embedded Markov chain. Due to the periodic nature of the probe-and-release algorithm, if observed at each time $t$, the process $\{(x(t),e(t));t\ge0\}$ is non-Markovian. However, when observed at the end of each round, the process $\{(x[n],e[n]);n\ge0\}$ is Markovian. In this and the next subsections, we will take advantage of this property and obtain convergence from the theory of Markov chains.



\subsubsection{Boundedness of $e_R$}
Define the Lyapunov function $V(e_R)=e_{R}^2$. Apparently, $\mathrm{E}[\vartheta_R]=R$ and ${\mathrm{Var}}[\vartheta_{R}]=\sigma_{\varepsilon}^2$. We can obtain
\begin{align*}
    &\mathrm{E}[e_{R}^2[n]|e_{R}[n-1]= e_{R}]\\
    =&\mathrm{E}\Big[\frac{1}{R^2}\big(\sum_{j=1}^{k}\lambda(1-\lambda)^{k-j} \vartheta_{R}^{(j)}+(1-\lambda)^k\hat{R}[n-1]-R\big)^2\Big]\\
    =&\mathrm{E}\Big[\big(\frac{1}{R}\sum_{j=1}^{k}\lambda(1-\lambda)^{k-j}(\vartheta_{R}^{(j)}-R)+(1-\lambda)^{k}e_{R}\big)^2\Big]\\
=&\frac{\lambda(1-(1-\lambda)^{2k})}{2-\lambda}\frac{\sigma_{\varepsilon}^2}{R^2}+(1-\lambda)^{2k}e_{R}^2.
\end{align*}
The drift of $e_{R}^2$ for the $n$th round is given by
\begin{align*}
    &\Delta(e_{R}^2)=\mathrm{E}[e_{R}^2[n]|e_{R}[n-1]= e_{R}]-e_{R}^2\\
    =&\frac{\lambda(1-(1-\lambda)^{2k})}{2-\lambda}\frac{\sigma_{\varepsilon}^2}{R^2}-(1-(1-\lambda)^{2k})e_{R}^2.
\end{align*}
Define $c_1=1-(1-\lambda)^{2k}>0$, $d_1=\frac{\lambda(1-(1-\lambda)^{2k})}{2-\lambda}\frac{\sigma_{\varepsilon}^2}{R^2}<\infty$, $f(e_R)=e_R^2$,
then
$\Delta(e_{R}^2)\leq -c_1f(e_R)+d_1$.
By the Foster-Lyapunov Comparison Theorem ~\cite[Theorem 4.3]{meyn1993stability}, for any initial condition $e_R[0]=e_R\in \mathbb{R}$, we have
\begin{equation*}
    \limsup _{t \rightarrow \infty} \frac{1}{t} \sum_{n=0}^{t} \mathrm{E}[f(e_R[n]) |e_R[0]=e_R] \leq \frac{d_1}{c_1}=\frac{\lambda\sigma_{\varepsilon}^2}{(2-\lambda)R^2}.
\end{equation*}
\subsubsection{Boundedness of $e_{\alpha}$}
The proof is similar to above.
We define the Lyapunov function  $V(e_\alpha)=e_\alpha^2$. Since $\vartheta_{\alpha}=\frac{F(t)-x_0^{\mathrm{clean}}}{x_0(t)-x_0^{\mathrm{clean}}}$, where $F(t)=\alpha(x_0(t)-x_0^{\mathrm{clean}})+x_0^{\mathrm{clean}}+\varepsilon(t)\frac{x_0(t)-x_0^{\mathrm{clean}}}{x_0^{\mathrm{c}}-x_0^{\mathrm{clean}}}$,
we have $\mathrm{E}[\vartheta_\alpha]=\mathrm{E}[\alpha+\frac{\varepsilon(t)}{x_0^{\mathrm{c}}-x_0^{\mathrm{clean}}}]=\alpha$ and ${\mathrm{Var}}[\vartheta_{\alpha}]=\frac{\sigma_{\varepsilon}^2}{{(x_0^{\mathrm{c}}-x_0^{\mathrm{clean}}})^2}$. 
Thus,
\begin{align*}
    &\Delta(e_{\alpha}^2)=\mathrm{E}[e_{\alpha}^2[n]|e_{\alpha}[n-1]= e_{\alpha}]-e_{\alpha}^2\\
    =&\frac{\lambda(1-(1-\lambda)^{2k})\sigma_{\varepsilon}^2}{(2-\lambda)\alpha^2 (x_0^{\mathrm{c}}-x_0^{\mathrm{clean}})^2}-(1-(1-\lambda)^{2k})e_{\alpha}^2.
\end{align*}
Define $d_2=\frac{\lambda(1-(1-\lambda)^{2k})\sigma_{\varepsilon}^2}{(2-\lambda)\alpha^2 (x_0^{\mathrm{c}}-x_0^{\mathrm{clean}})^2}<\infty$, $f(e_\alpha)=e_\alpha^2$, 
then 
$\Delta(e_{\alpha}^2)\leq -c_1f(e_\alpha)+d_2$.
Similarly, for any initial condition $e_\alpha[0]=e_\alpha\in \mathbb{R}$, we have
\begin{align*}
    \limsup _{t \rightarrow \infty} \frac{1}{t} \sum_{n=0}^{t} \mathrm{E}[f&(e_\alpha[n]) |e_\alpha[0]=e_\alpha] \leq \frac{d_2}{c_1}\\&=\frac{\lambda\sigma_{\varepsilon}^2}{(2-\lambda)\alpha^2(x_0^{\mathrm{c}}-x_0^{\mathrm{clean}})^2}.
\end{align*}

\subsubsection{Convergence of $e_{F_{\mathrm{max}}}$ and $e_{\varepsilon_{\mathrm{max}}}$}
We first prove that $\lim\limits_{n\to\infty} e_{F_{\mathrm{max}}}[n] = 0$ almost surely.
Since (\ref{equ_F}), the random sequence $\{\hat F_{\mathrm{max}}[n], n \in \mathbbm{Z}_{\geq 1}\}$ is a non-decreasing sequence and it has an upper bound $Q+\varepsilon_{\mathrm{max}}$. Hence, for every sample point $\omega \in \Omega$, $\lim\limits_{n\to \infty} \hat F_{\mathrm{max}}[n](\omega)$ exists.
Define 
\begin{align*}
    p(x_0)=\mathrm{Pr}\bigg\{&Q+\varepsilon_{\mathrm{max}}-f(x_0)-\\&\bigg(\mathbb{I}_{x_0\leq x_0^{\mathrm{c}}}\frac{x_0-x_0^{\mathrm{clean}}}{x_0^{\mathrm{c}}-x_0^{\mathrm{clean}}}+\mathbb{I}_{x_0> x_0^{\mathrm{c}}}\bigg)\varepsilon(t)\leq \eta \bigg\}
\end{align*}
for $x_0^{\mathrm{min}} \leq x_0 \leq x_0^{\mathrm{max}}$, where $\eta$ is a sufficiently small positive number. We assume $\mathrm{Pr}\left\{\varepsilon_{\mathrm{max}}-\eta\leq \varepsilon(t) \leq \varepsilon_{\mathrm{max}} \right\} >0$, thus, $p(x_0^{\mathrm{c}})>0$.
Recall that in episode 2, $x_0^{\mathrm{set}}$ is selected uniformly at random from $[x_0^{\mathrm{min}},x_0^{\mathrm{max}}]$, then the corresponding sample $x_0(t_j)$, where $1\leq j\leq k$, also follows this uniform distribution. The PDF is given by $\rho(x_0)=\frac{1}{x_0^{\mathrm{max}}-x_0^{\mathrm{min}}},x_0 \in [x_0^{\mathrm{min}},x_0^{\mathrm{max}}]$.  
Let
$P=\mathrm{Pr}\{Q+\varepsilon_{\mathrm{max}}-\vartheta_{F_{\mathrm{max}}}\leq \eta \}.$ Then, $P=\int_{x_0^{\mathrm{min}}}^{x_0^{\mathrm{max}}} \rho(x_0)p(x_0) d x_0>0.$
Due to cleaning operations, the sample values are i.i.d.. Then,
\begin{align*}
    \mathrm{Pr}\left\{Q+\varepsilon_{\mathrm{max}}-\hat F_{\mathrm{max}}[n]\leq \eta \right\}= 1-(1-P)^{nk},
\end{align*}
$$\lim_{n\to \infty} \mathrm{Pr}\big\{ \big| \hat F_{\mathrm{max}}[n] -Q-\varepsilon_{\mathrm{max}} \big| \leq \eta \big\} =1.$$
It indicates that $\hat F_{\mathrm{max}}[n] \stackrel{\tiny\mbox{in prob.}}{\longrightarrow} Q+\varepsilon_{\mathrm{max}}$.
Let $\{E_n; n \in \mathbbm{Z}_{\geq1} \}$ be a sequence of events, where $E_n$ refers to $|\hat F_{\mathrm{max}}[n] -Q-\varepsilon_{\mathrm{max}} \big| > \eta$. It satisfies
\begin{align*}
\sum_{n=1}^{\infty} \mathrm{Pr}\{E_n\}=\sum_{n=1}^{\infty} (1-P)^{nk} =\frac{(1-P)^k}{1-(1-P)^k} < \infty.
\end{align*}
One can check that $E_1\supset E_2 \supset \cdots \supset E_{\infty}$. According to the Borel-Cantelli Lemma\cite{durrett2019probability}, $\mathrm{Pr}\{\cap_{N=1}^{\infty} (\cup_{n=N}^{\infty} E_n) \}=0$, i.e., 
$$\forall \eta >0,\mathrm{Pr}\{|\lim_{n\to \infty} \hat F_{\mathrm{max}}[n] -Q-\varepsilon_{\mathrm{max}} \big| > \eta \}=0.$$
Hence,
$$\mathrm{Pr}\{\lim_{n\to \infty} \hat F_{\mathrm{max}}[n] =Q+\varepsilon_{\mathrm{max}} \}=1,$$
$$\lim\limits_{n\to\infty} e_{F_{\mathrm{max}}}[n] =\lim_{n\to \infty} \frac{\hat F_{\mathrm{max}}[n]-(Q+\varepsilon_{\mathrm{max}})}{Q+\varepsilon_{\mathrm{max}}}=0\quad a.s.$$

For the convergence of $e_ {\varepsilon_{\mathrm{max}}}$, 
the proofs are analogous to above and omitted here. It follows that
$$\mathrm{Pr}\{\lim_{n\to \infty} \hat \varepsilon_{\mathrm{max}}[n] =\varepsilon_{\mathrm{max}} \}=1,$$
\begin{align*}
    \lim_{n \to \infty} e_ {\varepsilon_{\mathrm{max}}}[n]=\lim_{n \to \infty}\frac{\hat\varepsilon_{\mathrm{max}}[n]-\varepsilon_{\mathrm{max}}}{\varepsilon_{\mathrm{max}}}=0 \quad a.s.
\end{align*}

\subsection{Boundedness of traffic states}\label{sub_traffic}
The number of time steps in a round is given by $$T=3k+k\sum_{i=1}^{3}T_{\mathrm{clean},i}+T_{\mathrm{release}}+T_{\mathrm{clean},4}.$$
Let $\Gamma_1=T(A_{\mathrm{max}}+B_{\mathrm{max}})$, $\Gamma_2=\frac{\Gamma_1}{\mu_1}$ and $c=\mu_2\kappa(\gamma)$, where $\mu_1<\frac{\Lambda}{-\delta_2}$, $0<\mu_2<\frac{1}{1-\mu_1}$.
We construct a compound Lyapunov function coupling the traffic states $x=[x_0,x_1,\dots,x_s,q]^{\mathrm{T}}$ and the normalized estimation errors $e=[e_{\alpha}, e_{F_\mathrm{max}}, e_{R}, e_{\varepsilon_\mathrm{max}}]^{\mathrm{T}}$ as 
$$W(x,e)=(\Vert x \Vert_1 +1) \Vert e \Vert_2^2+\beta \Vert x \Vert_1^2,$$
where $\Vert x \Vert_1 = \sum_{i=0}^{s}|x_i|+|q|$, $\Vert e \Vert_2^2=e_{\alpha}^2+e_{F_\mathrm{max}}^2+e_{R}^2+e_{\varepsilon_\mathrm{max}}^2$, 
\begin{equation}
\label{equ_beta}
    \beta = \frac{\lambda(1-(1-\lambda)^{2k})}{(2\lambda-4)\Gamma_2}\Big(\frac{\sigma_{\varepsilon}^2}{\alpha^2{(x_0^{\mathrm{c}}-x_0^{\mathrm{clean}}})^2}+\frac{\sigma_{\varepsilon}^2}{R^2}\Big)-\frac{c}{2\Gamma_2}.
\end{equation}
Let
$$c'=-\Delta(\Vert {e} \Vert_2^2)-2\beta\Delta(\Vert x \Vert_1),$$ 
\begin{align*}
    d'=\big(\Vert {e} \Vert_2^2 +\Delta(\Vert {e} \Vert_2^2)\big) \Delta (\Vert {x} \Vert_1) +\beta(\Delta(\Vert {x} \Vert_1))^2 + \Delta(\Vert {e} \Vert_2^2),
\end{align*} 
the drift of $W(x,e)$ for the $n$th round is expanded as
\begin{align*}
&\Delta W(x,e)\\=&\mathrm{E}[W(x[n],e[n])|x[n-1]=x,e[n-1]=e]-W(x,e)\\=&\mathrm{E}\Big[\big(\|x\|_1+\Delta(\Vert {x} \Vert_1)+1\big)\big(\Vert {e} \Vert_2^2 +\Delta(\Vert {e} \Vert_2^2)\big)\\&+\beta\big(\|x\|_1+\Delta(\Vert {x} \Vert_1)\big)^2\Big]-\big(\Vert x \Vert_1 +1\big) \Vert e \Vert_2^2-\beta \Vert x \Vert_1^2\\=&-c'\Vert x \Vert_1 +d'.
\end{align*}
\begin{enumerate}
\item We first consider the case that  $\Vert {e} \Vert_2^2 \geq  4\gamma$.
$\hat{\alpha}$, $\hat{R}$ are updated by (\ref{equ_alpha}), (\ref{equ_R}) and the update equations of $\hat{F}_{\mathrm{max}}$ and $\hat{\varepsilon}_{\mathrm{max}}$ can be rewritten as
$$\hat{F}_{\mathrm{max}}[n]=\hat{F}_{\mathrm{max}}[n-1]+\zeta,$$
$$\hat{\varepsilon}_{\mathrm{max}}[n]=\hat{\varepsilon}_{\mathrm{max}}[n-1]+\tau,$$
where $\zeta$ and $\tau$ are two random variables, $0 \leq \zeta\leq |e_{F_\mathrm{max}}[n-1]|(Q+\varepsilon_{\mathrm{max}})$ and $0 \leq \tau \leq |e_{\varepsilon_\mathrm{max}}[n-1]|\varepsilon_{\mathrm{max}}$. 
The mean drift of $\Vert e\Vert_2^2$ for the $n$th round is derived as
\begin{align*}
    &\Delta(\Vert e \Vert_2^2)=\mathrm{E}[\Vert e[n] \Vert_2^2|e[n-1]= e]-\Vert e \Vert_2^2\\
    =&\frac{\lambda(1-(1-\lambda)^{2k})}{2-\lambda}\Big(\frac{\sigma_{\varepsilon}^2}{\alpha^2{(x_0^{\mathrm{c}}-x_0^{\mathrm{clean}}})^2}+\frac{\sigma_{\varepsilon}^2}{R^2}\Big)\\
    -&(1-(1-\lambda)^{2k})(e_{\alpha}^2+e_{R}^2)\\
    +&\mathrm{E}\Big[\frac{2\zeta e_{F_{\mathrm{max}}}}{Q+\varepsilon_{\mathrm{max}}} +\frac{\zeta^2}{(Q+\varepsilon_{\mathrm{max}})^2}+\frac{2\tau e_{\varepsilon_{\mathrm{max}}}}{\varepsilon_{\mathrm{max}}} +\frac{\tau^2}{(\varepsilon_{\mathrm{max}})^2}\Big].
\end{align*}
Let $\zeta'=\frac{\zeta}{Q+\varepsilon_{\mathrm{max}}}$ and $\tau'=\frac{\tau}{\varepsilon_{\mathrm{max}}}$, then
\begin{align*}
    &\mathrm{E}\Big[\frac{2\zeta e_{F_{\mathrm{max}}}}{Q+\varepsilon_{\mathrm{max}}} +\frac{\zeta^2}{(Q+\varepsilon_{\mathrm{max}})^2}\Big]\\=&\int_{0}^{|e_{F_{\mathrm{max}}}|} (2\zeta' e_{F_{\mathrm{max}}}+\zeta'^2) \varrho(\zeta') d\zeta'\\<& \int_{\frac{|e_{F_{\mathrm{max}}}|}{4}}^{|e_{F_{\mathrm{max}}}|} (2\zeta' e_{F_{\mathrm{max}}}+\zeta'^2) \varrho(\zeta') d\zeta'  \\
    <& \big(2 \frac{|e_{F_{\mathrm{max}}}|}{4} e_{F_{\mathrm{max}}}+\frac{|e_{F_{\mathrm{max}}}|^2}{16} \big) \int_{\frac{|e_{F_{\mathrm{max}}}|}{4}}^{|e_{F_{\mathrm{max}}}|} \varrho(\zeta') d\zeta'\\
    =&-\frac{7}{16}   e_{F_{\mathrm{max}}}^2 p\big(|e_{F_{\mathrm{max}}}|(Q+\varepsilon_{\mathrm{max}})\big),
\end{align*}
where $\varrho(\zeta')$ is the PDF of $\zeta'$ and $p(\cdot)$ is given by (\ref{equ_p}). Similarly,
\begin{align*}
    \mathrm{E}\Big[\frac{2\tau e_{\varepsilon_{\mathrm{max}}}}{\varepsilon_{\mathrm{max}}} +\frac{\tau^2}{(\varepsilon_{\mathrm{max}})^2}\Big]< -\frac{7}{16}   e_{\varepsilon_{\mathrm{max}}}^2 p'(|e_{\varepsilon_{\mathrm{max}}}|{\varepsilon_{\mathrm{max}}}),
\end{align*}
where $p'(\cdot)$ is given by (\ref{equ_p'}).
We define $\kappa(\gamma)$ by (\ref{equ_Kappa}).
Since $p(\sqrt{\gamma}(Q+\varepsilon_{\mathrm{max}}))$ and $p'(\sqrt{\gamma} \varepsilon_{\mathrm{max}})$ are positive constants, $\kappa(\gamma)$ and $c$ are also positive constants. 
The condition $\Vert e \Vert_2^2 \geq 4\gamma$ implies $\mathrm{max}\{e_{\alpha}^2,e_{F_\mathrm{max}}^2,e_{R}^2,e_{\varepsilon_\mathrm{max}}^2\} \geq \gamma$, one can check that
\begin{align*}
    \kappa(\gamma)&<\big((1-(1-\lambda)^{2k})(e_{\alpha}^2+e_{R}^2)+\frac{7}{16}   e_{F_{\mathrm{max}}}^2 \\
    p&(|e_{F_{\mathrm{max}}}|(Q+\varepsilon_{\mathrm{max}}))+\frac{7}{16}   e_{\varepsilon_{\mathrm{max}}}^2 p'(|e_{\varepsilon_{\mathrm{max}}}|\varepsilon_{\mathrm{max}})\big)_{\mathrm{min}}\\
    &<\big((1-(1-\lambda)^{2k})(e_{\alpha}^2+e_{R}^2)-\mathrm{E}[2\zeta' e_{F_{\mathrm{max}}}+\zeta'^2]\\&\qquad-\mathrm{E}[2\tau' e_{\varepsilon_{\mathrm{max}}}+\tau'^2]\big)_{\mathrm{min}}.
\end{align*}
Considering the most conservative case (though unrealizable in practice) that the inflow is $A_{\mathrm{max}}+B_{\mathrm{max}}$ and the outflow is $0$ for each step in a round, we can obtain the mean drift of $\|x\|_1$ for the $n$th round 
\begin{align*}
\Delta(\Vert x\Vert_1)&=\mathrm{E}[\Vert x[n]\Vert_1|x[n-1]={x}]- \Vert {x} \Vert_1 \\ &\leq T(A_{\mathrm{max}}+B_{\mathrm{max}})=\Gamma_1.
\end{align*}
Define $\omega=-2\beta\Gamma_1-c$.
Since (\ref{equ_noise_var}),
there exists $\gamma$ such that
\begin{align*}
    \frac{(\frac{1}{1-\mu_1}-\mu_2)\kappa(\gamma)(2-\lambda)\alpha^2R^2(x_0^{\mathrm{c}}-x_0^{\mathrm{clean}})^2}{\lambda(1-(1-\lambda)^{2k})(R^2+\alpha^2(x_0^{\mathrm{c}}-x_0^{\mathrm{clean}})^2)}\geq \sigma_{\varepsilon}^2,
\end{align*}
then 
\begin{align*}
    \frac{\lambda(1-(1-\lambda)^{2k})}{2-\lambda}\Big(\frac{\sigma_{\varepsilon}^2}{\alpha^2{(x_0^{\mathrm{c}}-x_0^{\mathrm{clean}}})^2}+\frac{\sigma_{\varepsilon}^2}{R^2}\Big)\\\leq \Big(\frac{1}{1-\mu_1}-\mu_2\Big)\kappa(\gamma),
\end{align*}
\begin{align*}
    \Big(1-\frac{\Gamma_1}{\Gamma_2}\Big)\frac{\lambda(1-(1-\lambda)^{2k})}{2-\lambda}\Big(\frac{\sigma_{\varepsilon}^2}{\alpha^2{(x_0^{\mathrm{c}}-x_0^{\mathrm{clean}}})^2}+\frac{\sigma_{\varepsilon}^2}{R^2}\Big)\\-\kappa(\gamma)+(1-\frac{\Gamma_1}{\Gamma_2})c\leq 0,
\end{align*}
\begin{align*}
    \frac{\lambda(1-(1-\lambda)^{2k})}{2-\lambda}\Big(\frac{\sigma_{\varepsilon}^2}{\alpha^2{(x_0^{\mathrm{c}}-x_0^{\mathrm{clean}}})^2}+\frac{\sigma_{\varepsilon}^2}{R^2}\Big)-\kappa(\gamma)\\\leq-2\beta\Gamma_1-c.
\end{align*}
Since $\Delta(\Vert e\Vert_2^2)\leq \frac{\lambda(1-(1-\lambda)^{2k})}{2-\lambda}\Big(\frac{\sigma_{\varepsilon}^2}{\alpha^2{(x_0^{\mathrm{c}}-x_0^{\mathrm{clean}}})^2}+\frac{\sigma_{\varepsilon}^2}{R^2}\Big)-\kappa(\gamma)$, we can obtain
$$\Delta(\Vert e\Vert_2^2)\leq\omega<0,$$
$$c=-\omega-2\beta\Gamma_1\leq-\Delta(\Vert e\Vert_2^2)-2\beta\Delta(\Vert x\Vert_1)= c'.$$
Let $d_1'=(4+\omega)\Gamma_1+\beta \Gamma_1^2+\omega.$
Since $\beta>0$,$\Vert {e} \Vert_2^2<(\frac{0-\alpha}{\alpha})^2+(\frac{0-(Q+\varepsilon_{\mathrm{max}})}{Q+\varepsilon_{\mathrm{max}}})^2+(\frac{0-R}{R})^2+(\frac{0-\varepsilon_{\mathrm{max}}}{\varepsilon_{\mathrm{max}}})^2=4$, $\Delta(\Vert e\Vert_2^2)\leq\omega$ and $\Delta(\Vert x\Vert_1)\leq \Gamma_1$, one can verify that $d_1'\geq d'$.

\item Then we consider the second case that $\Vert e \Vert_2^2 < 4\gamma$ and $\|x\|_{1}\geq\|x\|_{1}^{\mathrm{th}}$, where $\|x\|_{1}^{\mathrm{th}}=TQ$.
Considering the most conservative mean outflow $\tilde{R}$ during $T_{\mathrm{release}}$ steps, we can obtain
\begin{align*}
    \Delta(\Vert {x} \Vert_1)<&(3k+k\sum_{i=1}^{3}T_{\mathrm{clean},i}+T_{\mathrm{clean},4})\\&(A_{\mathrm{max}}+B_{\mathrm{max}})-T_{\mathrm{release}}(\tilde{R}-\bar{A}-\bar{B}).
\end{align*}
According to (\ref{equ_T_rel}), 
\begin{align*}
    T_{\mathrm{release}}&\geq \frac{(1-\frac{1}{\mu_1})(3k+k\sum_{i=1}^{3}T_{\mathrm{clean},i}+T_{\mathrm{clean,4}})}{\frac{\delta_2}{\Lambda}+\frac{1}{\mu_1}}\\
    &\geq \frac{(1-\frac{1}{\mu_1})(3k+k\sum_{i=1}^{3}T_{\mathrm{clean},i}+T_{\mathrm{clean,4}})}{\frac{\tilde{R}-\bar{A}-\bar{B}}{A_{\mathrm{max}}+B_{\mathrm{max}}}+\frac{1}{\mu_1}}.
\end{align*}
Since $\mu_1<-\frac{\Lambda}{\delta_2}$, then $\frac{\tilde{R}-\bar{A}-\bar{B}}{A_{\mathrm{max}}+B_{\mathrm{max}}}+\frac{1}{\mu_1}\geq\frac{\delta_2}{\Lambda}+\frac{1}{\mu_1}>0$,
\begin{align*}
    T_{\mathrm{release}}(\frac{\tilde{R}-\bar{A}-\bar{B}}{A_{\mathrm{max}}+B_{\mathrm{max}}}+\frac{1}{\mu_1})\geq (1-\frac{1}{\mu_1})\\(3k+k\sum_{i=1}^{3}T_{\mathrm{clean},i}+T_{\mathrm{clean,4}}),
\end{align*}
\begin{align*}
    &T_{\mathrm{release}}(\tilde{R}-\bar{A}-\bar{B})\\\geq&(3k+k\sum_{i=1}^{3}T_{\mathrm{clean},i}+T_{\mathrm{clean,4}})(A_{\mathrm{max}}+B_{\mathrm{max}})\\&-\frac{1}{\mu_1}T(A_{\mathrm{max}}+B_{\mathrm{max}})\\
    =&(3k+k\sum_{i=1}^{3}T_{\mathrm{clean},i}+T_{\mathrm{clean,4}})(A_{\mathrm{max}}+B_{\mathrm{max}})-\Gamma_2.
\end{align*}
Thus, we can obtain $\Delta(\Vert {x} \Vert_1)<\Gamma_2<0.$
Let $\Gamma_3=\max\{(\bar{A}+\bar{B}-Q)T,-\|x\|_1\}$,
\begin{align*}
    d_2'=\beta\Gamma_3^2+\frac{\lambda(1-(1-\lambda)^{2k})}{2-\lambda}(\frac{\sigma_{\varepsilon}^2}{\alpha^2{(x_0^{\mathrm{c}}-x_0^{\mathrm{clean}}})^2}+\frac{\sigma_{\varepsilon}^2}{R^2}).
\end{align*}
Since (\ref{equ_beta}), $\Delta(\Vert e\Vert_2^2)\leq \frac{\lambda(1-(1-\lambda)^{2k})}{2-\lambda}(\frac{\sigma_{\varepsilon}^2}{\alpha^2{(x_0^{\mathrm{c}}-x_0^{\mathrm{clean}}})^2}+\frac{\sigma_{\varepsilon}^2}{R^2})$ and $\Gamma_3\leq\Delta(\Vert x\Vert_1)<\Gamma_2$, one can verify that 
\begin{align*}
    c=-&\frac{\lambda(1-(1-\lambda)^{2k})}{2-\lambda}\Big(\frac{\sigma_{\varepsilon}^2}{\alpha^2{(x_0^{\mathrm{c}}-x_0^{\mathrm{clean}}})^2}+\frac{\sigma_{\varepsilon}^2}{R^2}\Big)\\&\ \quad-2\beta\Gamma_2<-\Delta(\Vert e\Vert_2^2)-2\beta\Delta(\Vert x\Vert_1)= c'
\end{align*}
and $d_2'\geq d'$.

\item Finally, we consider the last case that $\Vert e \Vert_2^2 < 4\gamma$ and $\|x\|_{1}<\|x\|_{1}^{\mathrm{th}}$. 
Let
\begin{align*}
    d_3'=&\Big(c+\frac{\lambda(1-(1-\lambda)^{2k})}{2-\lambda}\big(\frac{\sigma_{\varepsilon}^2}{\alpha^2{(x_0^{\mathrm{c}}-x_0^{\mathrm{clean}}})^2}+\frac{\sigma_{\varepsilon}^2}{R^2}\big)\\&+2\beta T(\bar{A}+\bar{B})\Big)\|x\|_{1}^{\mathrm{th}}+
    4T(\bar{A}+\bar{B})\\&+\beta T^2\max\{(\bar{A}+\bar{B})^2,(\bar{A}+\bar{B}-Q)^2\}\\&+\frac{\lambda(1-(1-\lambda)^{2k})}{2-\lambda}\Big(\frac{\sigma_{\varepsilon}^2}{\alpha^2{(x_0^{\mathrm{c}}-x_0^{\mathrm{clean}}})^2}+\frac{\sigma_{\varepsilon}^2}{R^2}\Big).
\end{align*}
Since $T(\bar{A}+\bar{B}-Q)<\Delta(\Vert x\Vert_1)<T(\bar{A}+\bar{B})$, we can verify that
\begin{align*}
    &\Delta W(x, e)=-c\|x\|_{1}+\left(c-c'\right)\|x\|_{1}+d' \\
    \leq&-c\|x\|_{1}+\big(c+\Delta(\Vert {e} \Vert_2^2)+2\beta\Delta(\Vert x \Vert_1)\big)\|x\|_1^{\mathrm{th}}\\+&\big(\Vert {e} \Vert_2^2 +\Delta(\Vert {e} \Vert_2^2)\big) \Delta (\Vert {x} \Vert_1) +\beta(\Delta(\Vert {x} \Vert_1))^2 + \Delta(\Vert {e} \Vert_2^2)\\
    =&-c\|x\|_{1}+d_3'.
\end{align*}
\end{enumerate}
Let $d= \mathrm{max}\{d_1', d_2',d_3'\}$, one can check that $d<+\infty$. For the above three cases, 
$$\Delta W(x,e)=-c'\Vert x \Vert_1+d'\leq -c\Vert x \Vert_1+d.$$
According to Foster-Lyapunov Comparison Theorem \cite[Theorem 4.3]{meyn1993stability}, 
for any initial conditions $x[0]={x}\in \mathbbm{R}^{s+2}$ and $e[0]={e}\in \mathbbm{R}^{4}$, we have
$$\limsup_{t\to \infty} \frac{1}{t}\sum_{r=0}^t \mathrm{E}[\Vert x[r] \Vert_1|x[0]=x,e[0]=e]\leq \frac{d}{c},$$
that is, the number of vehicles on the road section under the probe-and-release algorithm is bounded on time-average.
\section{Implementation and validation}
\label{sec_simulation}
\begin{figure}[htp]
    \centering
    \includegraphics[width=\linewidth]{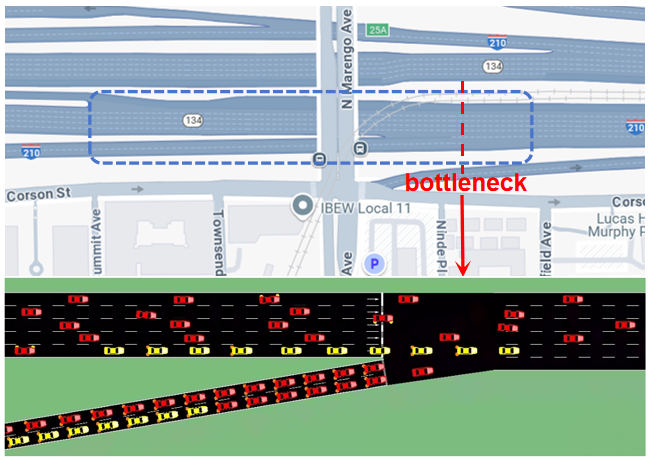}
    \caption{Interchange of I210 and SR134, and corresponding simulation scenario in SUMO with CAV platoons (yellow) and non-CAVs (red).}
    \label{fig_I210-134}
\end{figure}

In this section, we use real traffic data to conduct a set of simulation experiments to evaluate the probe-and-release control method. 
We carry out the simulation at the interchange of I210-E and SR134-E located in the Los Angles metropolitan area in California, United States (Fig.~\ref{fig_I210-134}). There are two lanes in upstream I210 and five lanes in upstream SR134, and the downstream section has a total of six lanes, thus forming a bottleneck at the interchange. We use the netedit in Simulation of Urban Mobility (SUMO) to draw the bottleneck. The total length of the road section is 1680 meters (from -1000 m \textendash\ 680 m) and the bottleneck is located at 680 m.

We first translate the control input $b_s(t)$ to speed instructions in Section \ref{sec_translate}. Then, we provide the setting, the estimation performance, and the control performance compared to the baseline and the state-of-the-art learning-based PPO method, under the stationary environment without pre-training in Section \ref{sec_simulation_setting}.
Finally, we compare the performance under a non-stationary environment in Section \ref{sec_simulation_comparison}.
All experiments are run on the GPU of NVIDIA GeForce RTX 4070 Ti.

\subsection{Translate $b_s(t)$ to speed instructions}
\label{sec_translate}
We now discuss how the probe-and-release control, especially the control input $b_s(t)$, can be translated to implementable instructions for platoons. These coordinations are enabled by modern V2I technology. The decision variable is the travel speed across the highway segment, which is equivalent to the scheduled arrival time at the bottleneck. We explicitly exclude lower-level control actions (e.g., longitudinal/lateral control) from consideration, assuming platoons possess underlying controllers capable of executing the speed instructions. 

\begin{figure}[htp]
    \centering
    \includegraphics[width=0.95\linewidth]{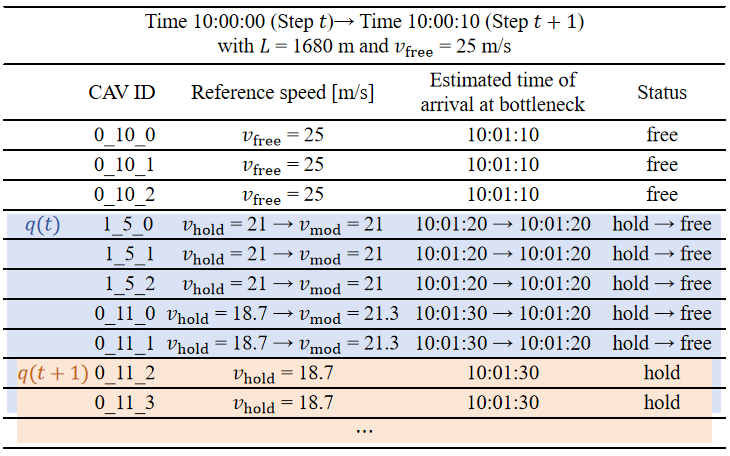}
    \caption{Traffic controller sends desired reference speed to each platoon. CAVs with status ``hold'' are in the virtual queue $q(t)$ and are assigned conservative speeds, computed according to real-time traffic conditions.}
    \label{fig_trans}
\end{figure}

Let $L$ (meters) be the length between the entrance and the bottleneck as in Fig. \ref{fig_abstract}, and $v_{\mathrm{free}}$ (m/s) be the free flow traverse speed (i.e., nominal speed) for the highway section. One step is $\Delta t$ seconds, then the free flow traverse time is given by $s \Delta t=\frac{L}{v_{\mathrm{free}}}$ seconds. Suppose that the current time step is $t$.
The first $b_{Bs}(t)$ vehicles in the CAV inflow (though the platoons may be split) are given the recommended speed $v_{\mathrm{free}}$, thus arriving at the bottleneck after $s$ steps. For the remaining $b_{Bq}(t)$ vehicles in the CAV flow, they will be postponed (i.e., join the virtual queue). Through considering the most conservative case that the future non-CAV inflows are $A_{\mathrm{max}}$, the recommended holding speeds for them can be given by $$v_{\mathrm{hold}}=\frac{L}{(s+\ell)\Delta t},\ \ell\in \mathbb{Z}_{\geq 1},$$
where the computation of $\ell$ is shown in Appendix~\ref{app_ell}.
In addition, the first $b_{qs}(t)$ vehicles in the virtual queue receive instructions to accelerate from the holding speed $v_\mathrm{hold}$ to the modified speed $$v_{\mathrm{mod}}=\frac{L-\Delta t(t-t_0)v_{\mathrm{hold}}}{s\Delta t},$$ where $t_0$ is the entry time step of the CAV. Note that for the first $b_{Bs}(t)$ vehicles in the CAV inflow, only one speed instruction $v_{\mathrm{free}}$ is sent when they enter the highway, and no more instructions are needed. However, for the remaining $b_{Bq}(t)$ vehicles in the CAV flow, two instructions are required: initial postponement with $v_\mathrm{hold}$ followed by acceleration to $v_\mathrm{mod}$. We use Fig. \ref{fig_trans} to illustrate the above process, and note that from step $t$ to $t+1$, with $b_{qs}(t)$ vehicles released and $b_{Bq}(t)$ vehicles joining, the vehicle sequence of the virtual queue is updated.
In practice, to prevent the virtual queue from blocking the upstream traffic, we let all queuing CAVs enter the same lane. 

\subsection{Stationary environment without pre-training}
\label{sec_simulation_setting}
In this set of experiments, we calibrate a stationary simulation environment with time-invariant parameters. The initial estimation is arbitrarily selected without prior knowledge.

\subsubsection{Simulation setup}
We acquired traffic data from the Freeway Performance Measurement System (PeMS) maintained by the California Department of Transportation, USA. 
We used data from March 1 to May 31, 2025 for the calibration.
We assume a 50\% penetration rate of CAVs.
Thus ground truth of the traffic demand is
$A_{\mathrm{max}}=B_{\max}=5.4$ veh/step and $\bar A=\bar B=3.6$ veh/step, where we take 10 seconds as a time step in simulation. Consequently, in free flow, it takes $s=7$ steps (i.e., 70 sec) for a vehicle to join the traffic queue at the bottleneck.

\begin{figure}[htp]
    \centering
    \includegraphics[width=0.95\linewidth]{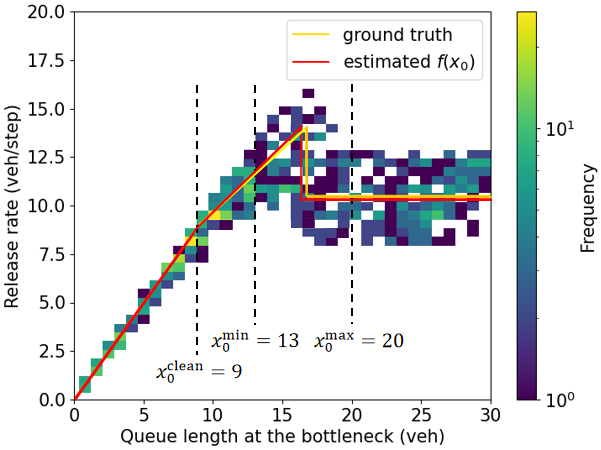}
    \caption{Data density scatter plot and flow function estimated by the probe-and-release algorithm.}
    \label{fig_density_scatter}
\end{figure}

With the calibrated environment, we simulated the flow out of the bottleneck in the face of stationary traffic demand and stationary capacity (subject to white noises).
Fig. \ref{fig_density_scatter}
illustrates the sample points $(x_0(t),F(t))$ for the flow function. The ground truth of parameters unknown to the probe-and-release algorithm are determined via the minimum square error method: $\alpha=0.65$/step, $R=10.5$ veh/step (i.e., 3780 veh/hr), $F_{\mathrm{max}}=16$ veh/step (i.e., 5760 veh/hr), $ \varepsilon_{\mathrm{max}}=2$ veh/step.
The variance of noise is determined to be $\sigma_{\varepsilon}^2=1.42$ (veh/step)$^2$. It is intuitive that the data points with high density (yellow ones and green ones) either exactly fall on the ground truth or closely around it.

We now verify that the simulation environment satisfies the prerequisites for Theorem~\ref{theo_stability}. First, $\tilde R$ is calculated to be 10.77 veh/step, which is greater than $\bar A+\bar B=7.2$ veh/step.
Second, $A_{\max}=5.4$ veh/step, which is smaller than $\min\{x_0^{\mathrm{clean}},R-\varepsilon_{\max}\}=8.5$ veh/step.
Third, there exists $\gamma=0.04$ such that the right hand side of \eqref{equ_noise_var} is $1.21$ (veh/step)$^2$, which is greater than $\sigma_{\varepsilon}^2=1.08$ (veh/step)$^2$.
Hence, the fluid queuing model with the above parameters is guaranteed to be stable under the probe-and-release algorithm. Note that $\tilde{R}>R$, which means the algorithm can stabilize the higher-demand system, thus to improve the overall throughput compared to the case without coordination.
The parameters of prior knowledge are
    $x_0^{\mathrm{clean}}=9$ veh,
    $x_0^{\mathrm{min}}=13$ veh, $x_0^{\mathrm{max}}=20$ veh,
    $\delta_1=3$ veh/step, 
    $\delta_2=3.5$ veh/step,
    $\Lambda=11$ veh/step,
    $\mu_1=-90$,
some of which are also indicated in Fig.~\ref{fig_density_scatter}.
The clean times and the release time can be computed from the prior knowledge:
$T_{\mathrm{clean},1}=2$ steps, $T_{\mathrm{clean},2}=4$ steps, $T_{\mathrm{clean},3}=7$ steps, $T_{\mathrm{clean},4}=51$ steps, $T_{\mathrm{release}}=327$ steps.

\label{sec_simulation_estimation results}

\subsubsection{Estimation performance} 
For initial estimates, $\hat F_{\mathrm{max}}[0]$ and $\hat\varepsilon_{\mathrm{max}}[0]$ should be set as $0$ veh/step, while $\hat\alpha[0]$ and $\hat R[0]$ can be initialized with random values. We use (\ref{equ_alpha})\textendash(\ref{equ_noise}) to estimate the four parameters, with learning rate $\lambda=0.08$ and sample rate $k=3$. The final probing results are $\hat{\alpha}=0.67$/step, $\hat{R}=10.4$ veh/step, $\hat{F}_{\mathrm{max}}=16$ veh/step, $\hat \varepsilon_{\mathrm{max}}=2$ veh/step, and the estimated flow function is shown in Fig. \ref{fig_density_scatter}. The normalized error vector (with respect to the ground truth) has a norm of $\|e\|_2^2=0.001$.

\begin{figure}[htp]
    \centering
    \includegraphics[width=0.9\linewidth]{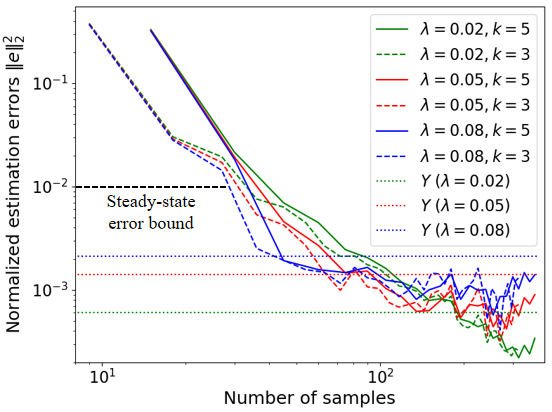}
    \caption{Simulation verifies the theoretical bound $Y$ for estimation error given by Theorem~\ref{theo_stability}.}
    \label{fig_err_trend}
\end{figure}

We also conduct sensitivity analysis for the design parameters $\lambda$ and $k$.
Fig. \ref{fig_err_trend}
shows the evolution of the normalized error with learning rate $\lambda=0.02,0.05,0.08$ and sample rate $k=3,5$, respectively.
Every combination of parameters are used in five repeated experiments.
%
The result implies that a higher learning rate $\lambda$ accelerates tracking but amplifies noise-induced fluctuations.
The sample rate $k$ turns out to have insignificant effects on the parameter estimation, especially in later stages, although a smaller sample rate yields a faster initial tracking.
During the initial stage, updates to $\hat F_{\mathrm{max}}$ and $\hat \varepsilon_{\mathrm{max}}$ drive the most substantial reductions in estimation errors, while during the later stage, updates to $\hat \alpha$ and $\hat R$ dominate the subtle fluctuations. 
The steady-state error bound is computed based on the definition of settle time ($\pm5\%$ error band): $\|[\pm0.05,-0.05,\pm0.05,-0.05]^{\mathrm{T}}\|_2^2=0.01$, and with no more than 30 samples, the normalized estimation errors can converge to the steady states.
Finally, the errors in all the cases are below the theoretical upper bound $Y$ in Theorem \ref{theo_stability}, as shown in Fig. \ref{fig_err_trend}.
In addition, the upper bound increases with the learning rate $\lambda$, which is also consistent with Theorem~\ref{theo_stability}.



\subsubsection{Control performance}
We now compare the control performance of the probe-and-release algorithm with two benchmarks:

\begin{enumerate}
    \item No coordination: all CAV platoons will travel according to the default driving behavior in SUMO.
    \item Proximal policy optimization (PPO): a representative reinforcement learning method very popular with transportation researchers. The state variable includes the vehicle density ahead and the ego vehicle's position, while the ego speed serves as the action. The reward function integrates a short-term factor of speed and a long-term factor of flow, promoting the traffic efficiency. We call the PPO algorithm from the Stable Baseline3 library with the following arguments: net architecture $[64,64]$, learning rate $0.002$, batch size 64 and discount factor 0.99. We use the coordination policy obtained by PPO as a proxy for the ground-truth optimal policy; the deviation between PPO and our algorithm is thus the approximate optimality gap.
\end{enumerate}

\begin{figure}[htp]
    \centering
    \includegraphics[width=0.8\linewidth]{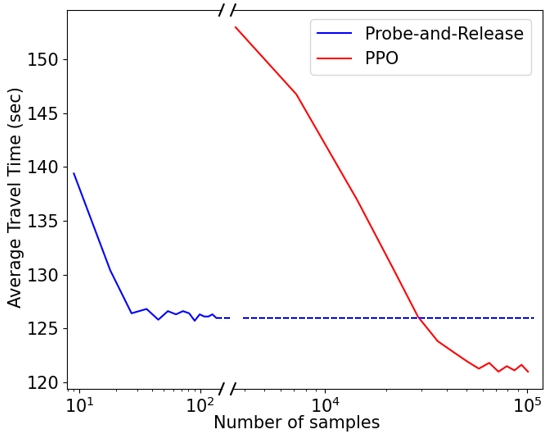}
    \caption{Our algorithm converges using merely 0.05\% of the samples required by PPO, with an insignificant optimality gap of 3.96\%.}
    \label{fig_comparison_stationary_v2}
\end{figure}

Fig. \ref{fig_comparison_stationary_v2} show that the average travel time converges to 126.1 sec for the probe-and-release method, and converges to 121.3 sec for PPO. Despite the minor gap in the average travel time, the probe-and-release method demonstrates a faster convergence speed with only 30 samples compared to PPO with around $6\times10^4$ samples, in the absence of pre-training. We will study the impact of pre-training in the next subsection.

\begin{figure*}[htp]
    \centering
    \includegraphics[width=\textwidth]{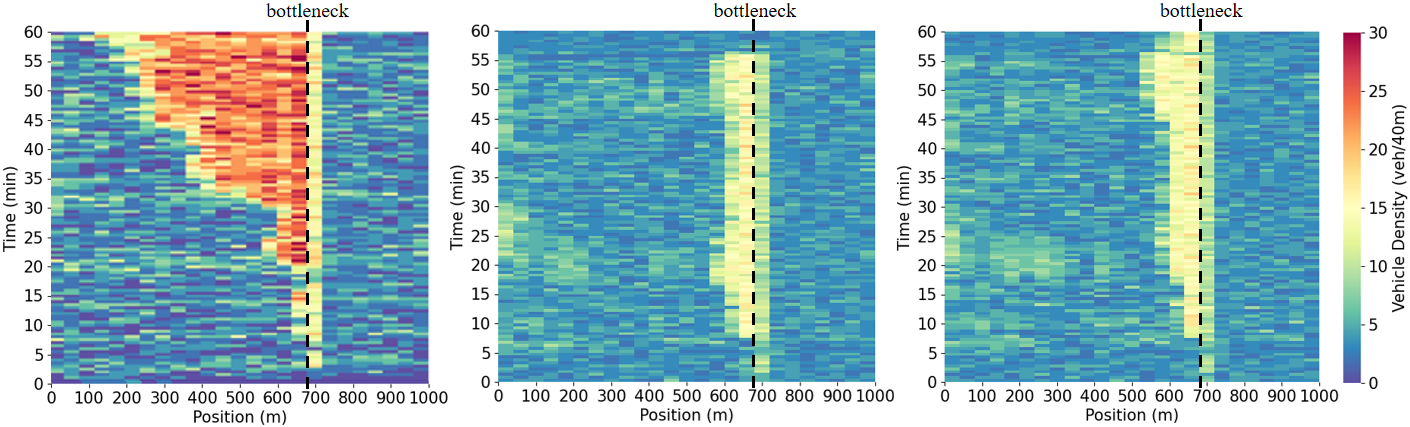}
    \caption{Traffic density heatmaps under no coordination (left), probe-and-release control (middle), and PPO (right). The demand and capacity are stationary with white noises.}
    \label{fig_heatmap}
\end{figure*}

Fig. \ref{fig_heatmap} shows the heatmaps of the road segment for the three methods. The bottleneck is located around 680 m. Without control, the vehicle queue accumulates rapidly upstream of the bottleneck, with very high vehicle density. Our method effectively mitigates congestion by regulating queued vehicles around the critical value $\hat x_0^{\mathrm{c}}=16.7 \ \mathrm{veh}$. Vehicles upstream of the bottleneck decelerate in advance, resulting in a more uniform density distribution compared to the no-coordination scenario. PPO also performs well by modulating upstream speeds, thus, smoothing the traffic flow upstream of the bottleneck.

\subsection{Non-stationary environment with pre-training}
\label{sec_simulation_comparison}

In this set of experiments, we consider a 24-hour simulation horizon with realistic variation of demand and capacity. The pre-trained controller obtained from the stationary environment is applied at 0:00. The objective of this subsection is to evaluate the capability of our algorithm to track and adapt to environmental changes.

\begin{figure}[htp]
    \centering
    \includegraphics[width=0.95\linewidth]{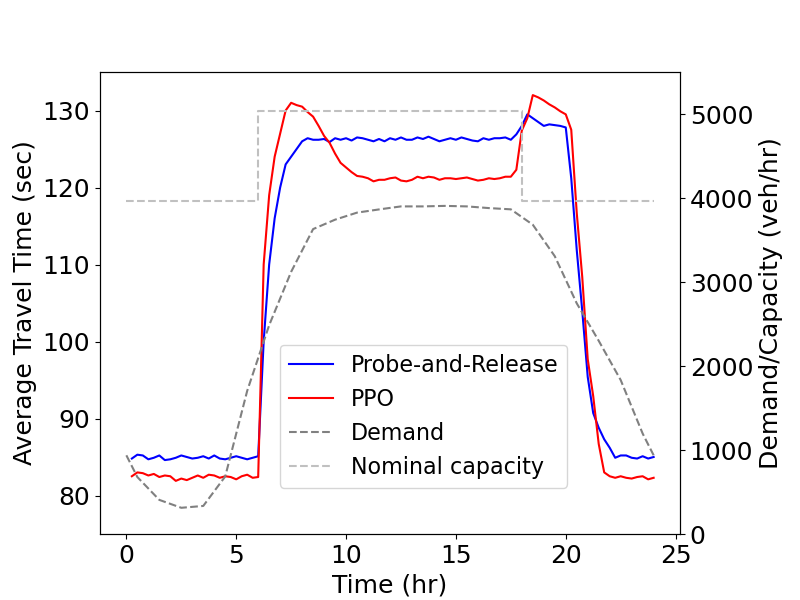}
    \caption{Probe-and-release algorithm turns out to track environmental changes faster than PPO, although PPO performs better after settling down. Note that the daytime is associated with a smaller capacity-to-demand margin and thus a higher average travel time.}
    \label{fig_comparison_changing}
\end{figure}

Fig.~\ref{fig_comparison_changing} shows the actual demand curve over one day and the capacity variation due to day-night shifts. In this setting, the demand increases from 5:00 to 8:00 and decreases from 19:00 to 22:00, and the day-night shift occurs at 6:00 and 18:00, with a higher capacity in the daytime. We will use these two curves as the demand and capacity for simulation. In order to track environmental changes (especially a possible decrease in $F_{\mathrm{max}}$ and $\varepsilon_{\mathrm{max}}$), $\hat F_{\mathrm{max}}$ and $\hat \varepsilon_{\mathrm{max}}$ are reset periodically. Here, we set the period as 12 hours, corresponding to day-night shifts. Our method tracks the model parameters and PPO updates the policy both in an online manner.

The comparison between our method and PPO is presented in Fig. \ref{fig_comparison_changing}.
Despite the higher nominal capacity during the day, the average travel time in the daytime is still longer than at night since the increment of traffic demands significantly larger than the increment of nominal capacity. For the baseline system, the average travel time is around 149.6 sec during the day and 86.0 sec during the night. Our method and PPO algorithm both perform much better than the no-coordination scenario, with PPO policy achieving a relatively smaller value in the convergent state, but the gap between them is small (only 1.9 sec during the night and 4.8 sec during the day).  However, transient mismatches between probing results/policies and real-world conditions at day-night shifts temporarily inflate the average travel time, most notably around 18:00. At two shift points, our method presents a smaller overshoot and a faster convergence speed compared to PPO (2.0 hr vs. 4.6 hr).


The sensing period for the probe-and-release control is 10 sec, corresponding to the length of a time step, while PPO requires sensing every second. Thus, our algorithm achieves comparable performance while utilizing only 10\% of the data volume required by PPO.

In conclusion, the performance comparison among the three methods is presented in Table \ref{table_perf_com}. The probe-and-release control method achieves comparable travel time performance, while demonstrating significantly reduced data requirements and substantially shorter training time compared to PPO. These are the key advantages of our new method.
\begin{table}[htp]
\centering
\caption{Performance comparison.}
\label{table_perf_com}
\begin{tabular}{m{2.7cm}<{\centering} m{1.3cm}<{\centering} m{1.8cm}<{\centering} m{1.3cm}<{\centering}}
\toprule
Metric & No coordination & Probe-and-Release & PPO \\
\midrule
Avg. Travel Time (Day) [sec] & 149.6 & 126.1 & 121.3 \\
\midrule
Avg. Travel Time (Night) [sec] & 86.0 & 84.7 & 82.8 \\
\midrule
Tracking Time (With Pre-training) [hr] & -- & 2.0 & 4.6 \\
\midrule
Sensing period [sec] &--& 10 & 1\\
\bottomrule
\end{tabular}
\end{table}
\section{Concluding remarks}
\label{sec_conclusion}
This paper studied inter-platoon coordination at highway bottlenecks with capacity breakdown in mixed-autonomy traffic.  Based on a fluid queuing model, we developed the probe-and-release control algorithm for simultaneous estimation of parameters and release of traffic under unknown and non-stationary environments. The results guaranteed bounded estimation errors and bounded traffic queues. Furthermore, the proposed algorithm exhibited excellent performance in comparison to PPO in simulation tests. Future research directions will include the extension to traffic networks, a more practical and sophisticated model for non-CAVs, and the generalization to generic nonlinear systems with non-monotonic dynamics.

\appendix
\label{sec_appendix}

\subsection{Proof of Proposition \ref{prp1}}
\noindent\textbf{``$\Rightarrow$'':}


For the first condition, we define the Lyapunov function as $V(x)=\|x\|_1^2$, where the virtual queue $q\equiv0$ with no coordination. Then the mean drift is given by $\Delta V(x)=\mathrm{E}[V(x(t+1))|x(t)=x]-V(x)$.
According to \cite[Theorem 4.3]{meyn1993stability}, let
\begin{align*}
    &\tilde{c}=2(R-\bar{A}-\bar{B})>0,\ f(x)=\|x\|_1,\\
    &\tilde{d}_1=\bar{A}^2+\sigma_{A}^2+\bar{B}^2+\sigma_{B}^2+2\bar{A}\bar{B}+R^2+\sigma_{\varepsilon}^2 < \infty,\\
    &\tilde{d}_2=\tilde{d}_1-R^2+Q^2+2R[x_0^{\mathrm{c}}+s(A_{\mathrm{max}}+B_{\mathrm{max}})]<\infty.
\end{align*}
When $x_0(t)>x_0^{\mathrm{c}}$, 
\begin{align*}
    \Delta V(x)=& \mathrm{E}[(x_0(t)+x_1(t)-R-\varepsilon(t)+x_2(t)+\cdots\\&+x_s(t)+A(t)+B(t))^2]-(x_0(t)+x_1(t)\\&+\cdots+x_s(t))^2 \\
    =& 2(\bar{A}+\bar{B}-R)(x_0(t)+x_1(t)+\cdots+x_s(t))\\&+\mathrm{E}[A(t)^2+B(t)^2+2A(t)B(t)-2(A(t)\\&+B(t))R+ R^2+\varepsilon(t)^2]\\
    \leq& -\tilde{c}f(x)+\tilde{d}_1.
\end{align*}
When $x_0(t)\leq x_0^{\mathrm{c}}$, let $\epsilon(t)=\frac{x_0(t)-x_0^{\mathrm{clean}}}{x_0^{\mathrm{c}}-x_0^{\mathrm{clean}}}\varepsilon(t)$. Since $\varepsilon(t)$ is independent of $\frac{x_0(t)-x_0^{\mathrm{clean}}}{x_0^{\mathrm{c}}-x_0^{\mathrm{clean}}}$, $\mathrm{E}[\epsilon(t)]=\mathrm{E}[\frac{x_0(t)-x_0^{\mathrm{clean}}}{x_0^{\mathrm{c}}-x_0^{\mathrm{clean}}}]\mathrm{E}[\varepsilon(t)]=0$. Also since $\frac{x_0(t)-x_0^{\mathrm{clean}}}{x_0^{\mathrm{c}}-x_0^{\mathrm{clean}}}\leq 1$, $\mathrm{Var}[\epsilon(t)]\leq \sigma_{\varepsilon}^2$. Thus,
\begin{equation*}
\begin{aligned}
    \Delta V(x)\leq & \mathrm{E}[(x_0(t)+x_1(t)-\alpha x_0(t)-\epsilon(t)+x_2(t)+\cdots\\&+x_s(t)+A(t)+B(t))^2]-(x_0(t)+x_1(t)\\&+\cdots+x_s(t))^2 \\
    =& 2(\bar{A}+\bar{B}-\alpha x_0(t))(x_0(t)+x_1(t)+\cdots+x_s(t))\\&+\mathrm{E}[A(t)^2+B(t)^2+2A(t)B(t)-2\alpha x_0(t)\\
    &(A(t)+B(t))+ \alpha^2 x_0(t)^2+\epsilon(t)^2]\\
    \leq & -\tilde{c}f(x)+\tilde{d}_2.
\end{aligned}
\end{equation*}
Then for any initial condition $x(0)=x \in\mathbb R^{s+2}$, we have
\begin{equation*}
\begin{aligned}
    \limsup_{t\to \infty} \frac{1}{t}\sum_{r=0}^t \mathrm{E}[f(x(r))|x(0)=x]\leq \frac{\max\{\tilde{d}_1,\tilde{d}_2\}}{\tilde{c}}
    =\frac{\tilde{d}_2}{\tilde{c}},
\end{aligned}
\end{equation*}
that is, the number of vehicles in the system is bounded on time-average if $\Bar{A}+\bar{B}<R$.

For the second condition, we assume $x_i(0)\leq A_{\mathrm{max}}+B_{\mathrm{max}}$ without loss of generality, where $1\leq i \leq s$.
Then,
\begin{align*}
    x_0(1)=&x_0(0)+x_1(0)-F(0)\\\leq& (x_0(0)-F(0))_{\mathrm{max}}+A_{\mathrm{max}}+B_{\mathrm{max}}\\=&x_0^{\mathrm{c}}-(Q-\varepsilon_{\mathrm{max}})+A_{\mathrm{max}}+B_{\mathrm{max}}\leq x_0^{\mathrm{c}}.
\end{align*}
By analogy, $x_0(t) \leq x_0^{\mathrm{c}}$ for $2\leq t\leq s$. In addition,
\begin{align*}
    x_0(s+1)=&x_0(s)+A(0)+B(0)-F(s)\\\leq& (x_0(s)-F(s))_{\mathrm{max}}+A_{\mathrm{max}}+B_{\mathrm{max}}\\=&x_0^{\mathrm{c}}-(Q-\varepsilon_{\mathrm{max}})+A_{\mathrm{max}}+B_{\mathrm{max}}\leq x_0^{\mathrm{c}}.
\end{align*}
By analogy, $x_0(t) \leq x_0^{\mathrm{c}}$ for $t\geq s+2$. In summary, 
$\forall t\geq 0,x_0(t) \leq x_0^{\mathrm{c}},$ 
and $\forall t \geq 0, x_i(t)\leq A_{\mathrm{max}}+B_{\mathrm{max}}\leq Q-\varepsilon_{\mathrm{max}}$, where $1\leq i \leq s$. Hence, the system is apparently stable.$\hfill\qedsymbol$

\noindent\textbf{``$\Leftarrow$'':}

Except for the aforementioned two conditions, other conditions can be classified into at least one of the following cases:

\begin{enumerate}
    \item $\bar{A}+\bar{B}\geq R$ and $x_0(0)>x_0^{\mathrm{c}}$,



    \item $\bar{A}+\bar{B}\geq R$ and $A_{\mathrm{max}}+B_{\mathrm{max}}>Q-\varepsilon_{\mathrm{max}}$.

\end{enumerate}

We consider a bounded non-negative test function
$$W(x_0)=1-e^{-\theta x_0},$$ where $\theta$ is a positive constant. Let $C=[0,x_0^{\mathrm{c}}]$, $\delta x_0=A(t)+B(t)-R-\varepsilon(t)$,
then $\mathrm{E}[\delta x_0]=\bar{A}+\bar{B}-R\geq 0$ and $\forall x_0\in C^{c}=(x_0^{\mathrm{c}},\infty)$,
 \begin{align*}
    \Delta W(x_0)=&\mathrm{E}[W(x_0(t+1))|x_0(t)=x_0]-W(x_0)\\=&1-\mathrm{E}[e^{-\theta(x_0+\delta x_0)}]-(1-e^{-\theta x_0})\\
    =&e^{-\theta x_0}(1-\mathrm{E}[e^{- \theta (\delta x_0)}])
    =e^{-\theta x_0}(1-m_{\delta x_0}(-\theta)),
 \end{align*}
where $m_{\delta x_0}(-\theta)$ is the Moment Generating Function of the random variabe $\delta x_0$, and it can be expanded as follows:
$$1-m_{\delta x_0}(-\theta)=\theta \mathrm{E}[\delta x_0]+o(\theta).$$
Then there exists $\theta>0$ such that $\Delta W(x_0)\geq 0$.
Thus, we can apply the drift criterion for transience \cite[Theorem 8.0.2]{meyn2012markov} and conclude that the component $x_0$ of the state vector is transient, therefore, the stochastic process is transient and the whole system is unstable.$\hfill\qedsymbol$

\subsection{Computation of auxiliary quantities}

The following technical definitions are used in the theoretical analysis.
First, let 
\begin{equation}
\begin{aligned}
\label{equ_p}
    &p(\chi)=1-\Big(1-\int_{x_0^\mathrm{min}}^{x_0^{\mathrm{c}}}\rho(x_0)  \int_{\varepsilon_{\mathrm{max}}-\frac{3\chi}{4}+\alpha(x_0^{\mathrm{c}}-x_0)}^{\varepsilon_{\mathrm{max}}}\varphi(\varepsilon) d\varepsilon dx_0 \\
    &\qquad\qquad\qquad-\int_{x_0^{\mathrm{c}}}^{x_0^{\mathrm{max}}}\rho(x_0)\int_{\varepsilon_{\mathrm{max}}-\frac{3\chi}{4}+Q-R}^{\varepsilon_{\mathrm{max}}}\varphi(\varepsilon) d\varepsilon dx_0\Big)^k
\end{aligned}
\end{equation}
\begin{equation}
\begin{aligned}
\label{equ_p'}
    p'(\psi) =1-\Big(1-\int_{\varepsilon_{\mathrm{max}}-\frac{\psi}{2}}^{\varepsilon_{\mathrm{max}}} \varphi(\varepsilon) d\varepsilon-\int_{-\varepsilon_{\mathrm{max}}}^{-\varepsilon_{\mathrm{max}}+\frac{\psi}{2}} \varphi(\varepsilon) d\varepsilon\Big)^{k},
\end{aligned}
\end{equation}
where $\rho$ is the uniform PDF over $[x_0^{\min},x_0^{\max}]$ and $\varphi$ is the PDF of noise $\varepsilon$.

Next, we present the computation of $\tilde{R}$ via the formula
\begin{align*}
    \tilde{R}=\frac{1}{M}\sum_{m=1}^{M}g_{m,m-1}.
\end{align*}
%
%
%
For $1\leq m\leq s$, $\upsilon_{m}=\upsilon_{m-1}+Q-f(\upsilon_{m-1})-\varepsilon_{m-1}$, where $\upsilon_{0}=x_0^{\mathrm{clean}}+\frac{Q-x_0^{\mathrm{clean}}}{(1-2\sqrt{\gamma})\alpha}$.
Define
$$g_{m,0} = \int_{-\varepsilon_{\mathrm{max}}}^{\varepsilon_{\mathrm{max}}} \varphi(\varepsilon_{m-1}) f(\upsilon_{m}) d\varepsilon_{m-1},$$
$$g_{m,i} = \int_{-\varepsilon_{\mathrm{max}}}^{\varepsilon_{\mathrm{max}}} \varphi(\varepsilon_{m-1-i}) g_{m,i-1} d \varepsilon_{m-1-i},$$
where $\ 1 \leq i \leq m-1$.
For $s+1 \leq m\leq M$, $\upsilon_{m}=\upsilon_{m-1}+A_{m-s-1}+u_{m-s-1}-f(\upsilon_{m-1})-\varepsilon_{m-1}$, where $u_{m-s-1}=b_s(m-s-1)$. Define 
\begin{align*}
    g_{m,0} = \int_{0}^{A_{\mathrm{max}}}  \int_{0}^{B_{\mathrm{max}}} \int_{-\varepsilon_{\mathrm{max}}}^{\varepsilon_{\mathrm{max}}} \varphi(\varepsilon_{m-1}) f(\upsilon_{m}) d\varepsilon_{m-1} \\d\Phi_{B}(B_{m-s-1}) d\Phi_{A}(A_{m-s-1}),
\end{align*}
\begin{align*}
  g_{m,i} = \int_{0}^{A_{\mathrm{max}}}  \int_{0}^{B_{\mathrm{max}}} \int_{-\varepsilon_{\mathrm{max}}}^{\varepsilon_{\mathrm{max}}} \varphi(\varepsilon_{m-1-i}) g_{m,i-1} d \varepsilon_{m-1-i} \\d\Phi_{B}(B_{m-s-1-i}) d\Phi_{A}(A_{m-s-1-i}),\ 1\leq i \leq m-s-1,  
\end{align*}
$$g_{m,i} = \int_{-\varepsilon_{\mathrm{max}}}^{\varepsilon_{\mathrm{max}}} \varphi(\varepsilon_{m-1-i}) g_{m,i-1} d \varepsilon_{m-1-i},\ m-s \leq i \leq m-1,$$
where $\int d\Phi_{A}$ and $\int d\Phi_{B}$ are the Stieltjes integrals of the demand $A$ and $B$, respectively. 
Thus, $g_{m,m-1},1\leq m \leq M$ are obtained in a recursive manner. One can verify that $g_{m,m-1}$ is a constant, thus, $\tilde{R}$ is a constant. One can also verify that $g_{m,m-1}= \mathrm{E}[f(x_0(m))|x(0)=\xi_0], 1\leq m \leq M,$ 
then $\tilde{R}=\frac{1}{M}\sum_{m=1}^{M}\mathrm{E}[f(x_0(m))|x(0)=\xi_0]$, which is the mean outflow based on the initial condition $x(0)=\xi_0$. In fact, $\tilde{R}$ is the most conservative approximation of the mean outflow under small estimation errors $4\gamma$. If there is no noise, then the estimation error vanishes and $\tilde{R}$ can be equal to $Q$. In other words, despite the probe-and-release algorithm, the existence of noise leads to the mean outflow dropping from $Q$ to $\tilde{R}$.

\subsection{Detailed instructions for the virtual queue}
\label{app_ell}
We assume the virtual queue be composed of virtual $x_{s+\ell}(t), \ell \geq 1$. Vehicles in virtual $x_{s+\ell}(t)$ will spend $s+\ell$ steps entering the traffic queue at the bottleneck. Note that the instructions for the virtual queue aim at the delayed $b_{Bq}(t)$ CAVs and have no effect on the control input $b_s(t)$.

The current step is $t$. Considering the most conservative case that the future non-CAV demands are $A_{\mathrm{max}}$, the number of CAVs assigned to virtual $x_{s+\ell}(t+1), \ell \geq 1$ is given by $$b_{B \to s+\ell}(t)=\mathrm{min}\{b_{Bq}(t)-\sum_{i=1}^{\ell-1}b_{B\to s+i}(t), (b_{B \to s+\ell}^{\ast}(t))_{+}\},$$
and the recommended speed for these $b_{B \to s+\ell}(t)$ CAVs will be $v_{\mathrm{hold}}=\frac{L}{(s+\ell)\Delta t}$, where
\begin{align*}
    b_{B\to s+\ell}^{\ast}(t)=\hat x_0^{\mathrm{c}}[n]- x_0^{\mathrm{pred}}(t+s+\ell)+\\\hat{f}^{[n]}(x_0^{\mathrm{pred}}(t+s+\ell))-A_{\mathrm{max}}-\varepsilon_{\mathrm{max}},
\end{align*}
and $x_0^{\mathrm{pred}}(t+s+\ell)$ satisfies the following iteration:
\begin{align*}
    x_0^{\mathrm{pred}}(t+s+\ell)=&x_0^{\mathrm{pred}}(t+s+\ell-1)+A_{\mathrm{max}}+\varepsilon_{\mathrm{max}}\\&+x_{s+\ell}(t)-\hat{f}^{[n]}(x_0^{\mathrm{pred}}(t+s+\ell-1))
\end{align*}
for $\ell \geq 2$, and
$x_0^{\mathrm{pred}}(t+s+1)=x_0^{\mathrm{pred}}(t+s)+A_{\mathrm{max}}+\varepsilon_{\mathrm{max}}+b_s(t)-\hat{f}^{[n]}(x_0^{\mathrm{pred}}(t+s))$, with
${x}_0^{\mathrm{pred}}(t+s)$ given in (\ref{equ_x0_pred}).

\section*{References}{
}
\vspace{-1.3\baselineskip}  
\bibliographystyle{IEEEtran}
\bibliography{References}  

\begin{IEEEbiography}
[{\includegraphics[width=1in,height=1.25in,clip,keepaspectratio]{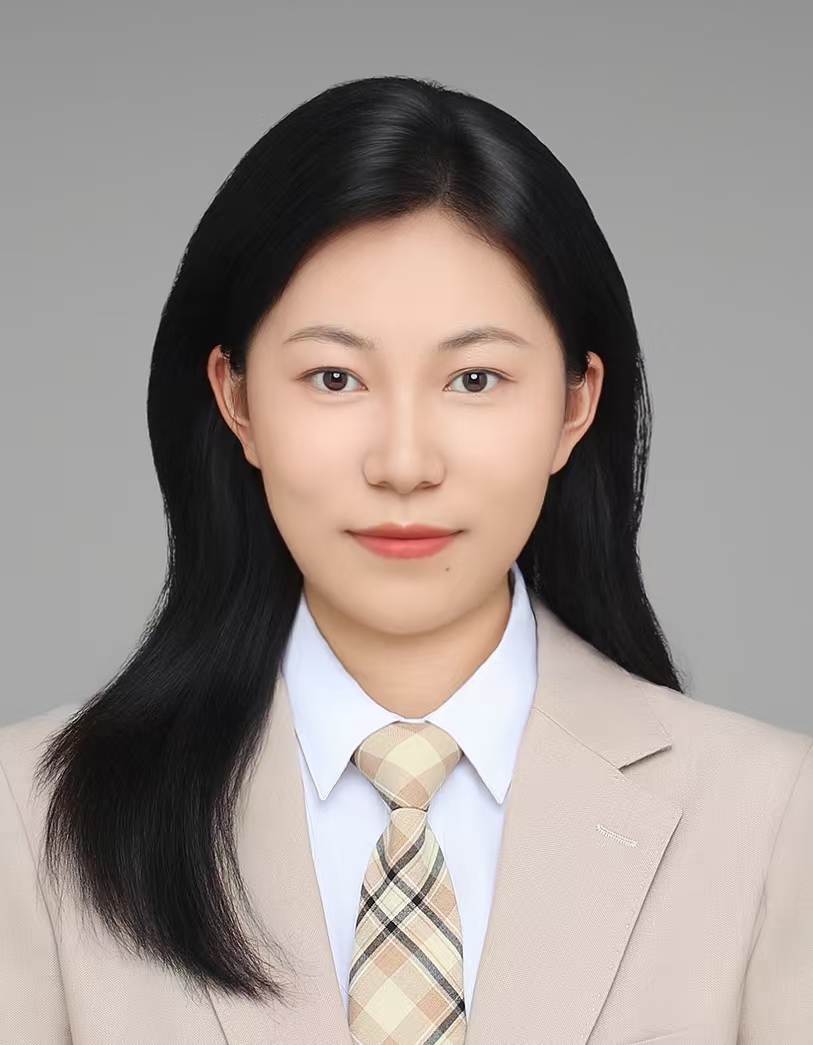}}]
{Yi Gao}
is a PhD student majoring in Control
Science and Engineering at the Global College,
Shanghai Jiao Tong University (SJTU), China. She
received her B.Eng. in Automation from Shandong University, China in 2022. She is interested in the control and optimization
for smart and connected transportation systems, with a focus on vehicle platooning and V2X coordination.
\end{IEEEbiography}

\begin{IEEEbiography}
[{\includegraphics[width=1in,height=1.25in,clip,keepaspectratio]{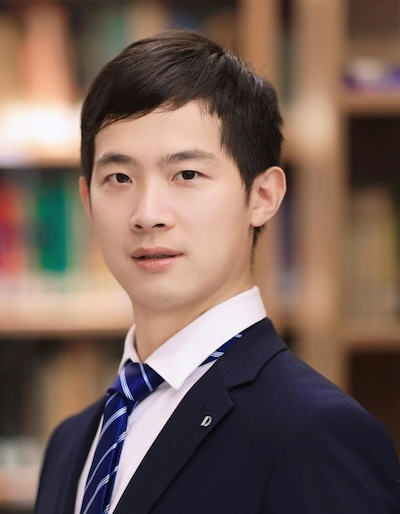}}]
{Xi Xiong}
received his bachelor's degree in Automotive Engineering from Jilin University, China, his master's degree in Mechanical Engineering from Tsinghua University, China, and his Ph.D. in Transportation Engineering from New York University, USA. He is currently a Research Professor of Transportation Engineering at Tongji University, Shanghai, China, and a Visiting Researcher at the Department of Engineering Science, University of Oxford, UK. Prior to joining Tongji, he was a Postdoctoral Fellow at the Harvard Kennedy School, USA. His research lies at the intersection of control, optimization, and machine learning, with a focus on enabling transformative change in societal systems.
\end{IEEEbiography}

\begin{IEEEbiography}
[{\includegraphics[width=1in,height=1.25in,clip,keepaspectratio]{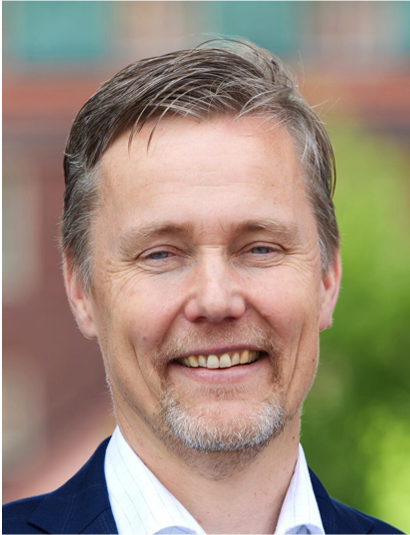}}]
{Karl H. Johansson} (Fellow, IEEE)
is Swedish Research Council Distinguished Professor in Electrical Engineering and Computer Science at KTH Royal Institute of Technology in Sweden and Founding Director of Digital Futures. He earned his MSc degree in Electrical Engineering and PhD in Automatic Control from Lund University. He has held visiting positions at UC Berkeley, Caltech, NTU and other institutions. His research interests focus on networked control systems and cyber-physical systems with applications in transportation, energy, and automation networks. 
He has also received the triennial IFAC Young Author Prize, IEEE CSS Distinguished Lecturer, IFAC Outstanding Service Award, and IEEE CSS Hendrik W. Bode Lecture Prize. 
His extensive service to the academic community includes being President of the European Control Association, IEEE CSS Vice President Diversity, Outreach and Development, and Member of IEEE CSS Board of Governors and IFAC Council. 
He has served on the editorial boards of Automatica, IEEE TAC, IEEE TCNS and many other journals. He has also been a member of the Swedish Scientific Council for Natural Sciences and Engineering Sciences. He is Fellow of the Royal Swedish Academy of Engineering Sciences.
\end{IEEEbiography}

\begin{IEEEbiography}[{\includegraphics[width=1in,height=1.25in,clip,keepaspectratio]{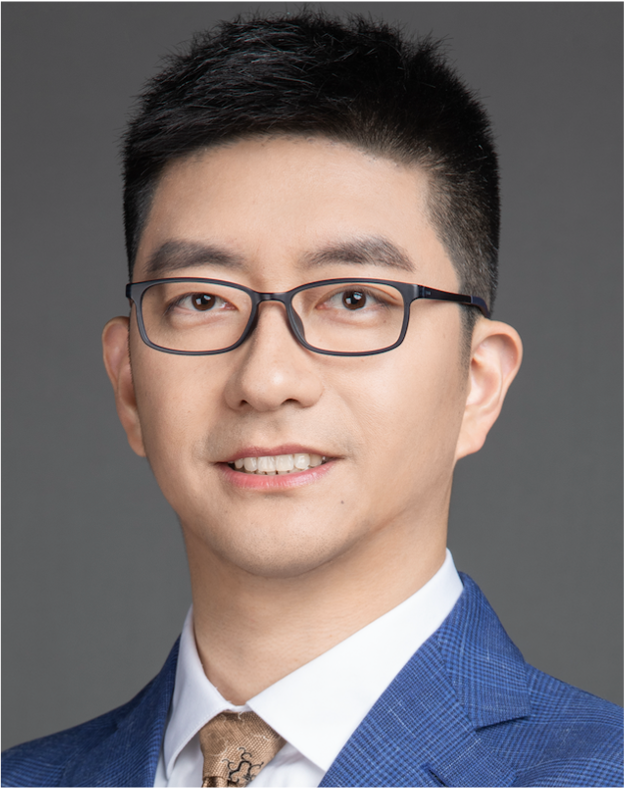}}]
{Li Jin} (Senior Member, IEEE)
is Associate Professor (2022--present) and was Assistant Professor (2021-2022) of Electrical and Computer Engineering at Shanghai Jiao Tong University (SJTU), China. He was Assistant Professor (2018--2020) at the Tandon School of Engineering, New York University, USA. He received his B.Eng. from SJTU in 2011, M.S. from Purdue University, USA in 2012, and Ph.D. from the Massachusetts Institute of Technology, USA in 2018. 
He is interested in connected and autonomous vehicles, network system control, and cyber-physical security.
\end{IEEEbiography}
\end{document}